\documentclass[epj,fleqn]{svjour}
\usepackage{pdfpages}
\usepackage[english]{babel}%
\usepackage{amsmath,amssymb}%
\usepackage{leftidx}%
\usepackage{graphicx}%
\usepackage[normal]{caption}
\usepackage{grffile}%
\usepackage{ifthen}%
\usepackage[squaren]{SIunits}%
\usepackage{xcolor}%
\usepackage{hyperref}%
\usepackage{dcolumn}
\usepackage{bm}
\usepackage[numbers,sort&compress]{natbib}
\usepackage{url}

\newcommand{\cf}{{cf.\ }}                            
\newcommand{\Tconst}{\mathrm{const}}                      
\newcommand{\eq}{eq.}                                     
\newcommand{\Eq}{Eq.}                                     
\newcommand{\eqs}{eqs.}                                   
\newcommand{\file}[1]{\nolinkurl{#1}}                           
\newcommand{\fig}{fig.}                                   
\newcommand{\figs}{figs.}                                 
\newcommand{\ie}{{i.\,e.\ }}                         
\newcommand{\eg}{{e.\,g.\ }}                         

\newcommand{\via}{{via}\ }                             

\newcommand{\treref}[2]{[notes p.~#1 %
  \ifthenelse{\equal{#2}{}}{} {eq.~(#2)}%
]}                                                  
\newcommand{\productspace}{\,}                            

\newcommand{\eqspace}{\;}                              
\newcommand{\Li}[2]{\mathrm{Li}_{#1}(#2)}                 
\newcommand{\LangevinFunction}{\mathcal L}                
\newcommand{\matsc}[3]{#1 
  \ifthenelse{\equal{#2}{}}{}{^{(#2)}}%
  \ifthenelse{\equal{#3}{}}{}{_{#3}}%
}%
\newcommand{\RiemannZeta}[1]{\zeta_{#1}}                  
\newcommand{\HeavisideTheta}[1]{\Theta({#1})}                  
\newcommand{\vecsc}[3]{#1 
  \ifthenelse{\equal{#2}{}}{}{^{(#2)}}%
  \ifthenelse{\equal{#3}{}}{}{_{#3}}%
}%
\addunit{\molar}{M}                                       
\addunit{\calory}{cal}                                    

\newcommand{\drm}{\mathrm{d}}                  
\newcommand{\e}{\mathrm{e}}                               
\newcommand{\kB}{\mathrm{k_B}}                            
\newcommand{\kBT}{\mathrm{k_B}T}                          
\newcommand{\timesspace}{\ }                   

\newcommand{\Rbp}[2]{(#1,#2)}                             
\newcommand{\Rc}{c}                                       
\newcommand{\Rcs}{\Rc^{*}}                                
\newcommand{\RcDNA}{{\hat\Rc}}                     
\newcommand{\RcDNAeff}{{\hat{\Rc}_{\mathrm{eff}}}}                     
\newcommand{\Reps}{\varepsilon}                 
\newcommand{\RF}{F}                                       
\newcommand{\RFc}{\RF_{\mathrm{c}}}                       
\newcommand{\Rfm}{\tau}                                   
\newcommand{\RG}{\mathcal G}                              
\newcommand{\RGG}{\mathcal G}                              
\newcommand{\Rg}{ g}                                      

\newcommand{\RGhinit}{\RG_{\mathrm{h}}^{\mathrm{init}}}   
\newcommand{\Rgh}{\Rg_{\mathrm{h}}}                
\newcommand{\RghHB}{\Rgh^{\mathrm{hb}}}                
\newcommand{\Rghstack}{\Rgh^{\mathrm{stack}}}                
\newcommand{\Rghi}{\Rgh^{\mathrm{i}}}                
\newcommand{\RGlentropy}{\RG_{\mathrm{l}}^{\mathrm{conf}}} 
\newcommand{\Rh}{h}                                        
\newcommand{\Rheatcap}{C}                       
\newcommand{\Rheatcapexponent}{\chi}          
\newcommand{\RheatcapexponentT}{\Rheatcapexponent} 
\newcommand{\Rk}{\kappa}          
\newcommand{\Rkb}{\kappa_{\mathrm{b}}} 
\newcommand{\Rkp}{\kappa_{\mathrm{p}}} 
\newcommand{\Rkc}{\kappa_{\mathrm{c}}} 
\newcommand{\RKuhnss}{b_{\mathrm{ss}}}                    
\newcommand{\RM}{M}                         
\newcommand{\Rm}{m}                         
\newcommand{\Rmu}{\mu}                      
\newcommand{\Rmud}{\mu_{\mathrm{d}}}        
\newcommand{\RN}{N}                         
\newcommand{\RPTn}{n}                       
\newcommand{\Rnp}{\theta}                   
\newcommand{\Rnpexponent}{\lambda}          
\newcommand{\RnpexponentT}{\Rnpexponent_{\RT}}          
\newcommand{\RnpexponentF}{\Rnpexponent_{\RF}}          
\newcommand{\RPG}{\Phi}                     
\newcommand{\Rs}{s}                                       
\newcommand{\Rsc}{\Rs_{\mathrm{c}}}      
\newcommand{\Rvf}[1]{v_{\mathrm{f}}(#1)}                  
\newcommand{\RT}{T}                                       
\newcommand{\RTc}{\RT_{\mathrm{c}}}                       
\newcommand{\RTmelt}{\RT_{\mathrm{m}}}                    
\newcommand{\RQH}{Q}               
\newcommand{\RQHt}{\tilde Q}   
\newcommand{\RQG}{Z}              
\newcommand{\RQGDNA}{\RQG^{\mathrm{DNA}}}        
\newcommand{\Rw}{w}               
\newcommand{\Rwinter}{\hat{w}}               
\newcommand{\Rwc}{\Rw_{\mathrm{c}}} 
\newcommand{\RwcPS}{\hat{\Rw}_{\mathrm{c}}} 
\newcommand{\RZG}{\mathcal Z}     
\newcommand{\RZGDNA}{\RZG^{\sqrt{\mathrm{DNA}}}}
\newcommand{\RZGb}{\RZG_{\mathrm{b}}}  
\newcommand{\Rz}{z}                    
\newcommand{\Rzd}{\Rz_{\mathrm{d}}}    
\newcommand{\Rzb}{\Rz_{\mathrm{b}}}    
\newcommand{\Rzp}{\Rz_{\mathrm{p}}}    
\newcommand{\Rzc}{\Rz_{\mathrm{c}}}    

\newcommand{\RzbPS}{\hat{\Rz}_{\mathrm{b}}}    
\newcommand{\RzpPS}{\hat{\Rz}_{\mathrm{p}}}    

\newcommand{\RK}{K}                                       
\newcommand{\RKb}{\RK_{\mathrm{b}}}
\newcommand{\RKp}{\RK_{\mathrm{p}}}
\newcommand{\RKcT}{\RK_{\mathrm{c,\RT}}}
\newcommand{\RKcF}{\RK_{\mathrm{c,\RF}}}

\newcommand{\DBstate}{B}
\newcommand{\DMstate}{M}

\newcommand{\DZ}{{\mathcal{Z}}}%
\newcommand{\DZB}{\hat{\DZ}_\mathrm{\DBstate}}%
\newcommand{\DZM}{\hat{\DZ}_\mathrm{\DMstate}}%
\newcommand{\DZPS}{\hat{\DZ}}

\newcommand{\subref}[2]{(#2)}
\newcommand{\subfloat}[1]{}
\begin{document}
\onecolumn

\title{Secondary structure formation of homopolymeric single-stranded
  nucleic acids including force and loop entropy: implications for DNA
  hybridization} \titlerunning{Secondary structures of single-stranded
  nucleic acids}


\author{Thomas R. Einert\inst{1} \and Henri Orland\inst{2} \and Roland
  R. Netz\inst{1,3}} \institute{Physik Department, Technische
  Universit\"at M\"unchen, James-Franck-Stra\ss e, 85748 Garching,
  Germany, Tel.: +49-89-28914337, Fax: +49-89-28914642,
  \email{einert@ph.tum.de} \and Institut de Physique Th\'eorique, CEA
  Saclay, 91191 Gif-sur-Yvette Cedex, France \and
  Fachbereich Physik, Freie Universit\"at Berlin, Arnimallee 14, 14195 Berlin, Germany
}
  
\abstract{ Loops are essential secondary structure elements in folded
  DNA and RNA molecules and proliferate close to the melting
  transition. Using a theory for nucleic acid secondary structures
  that accounts for the logarithmic entropy~$-\Rc\ln\Rm$ for a loop of
  length $\Rm$, we study homopolymeric single-stranded nucleic acid
  chains under external force and varying temperature.  In the
  thermodynamic limit of a long strand, the chain displays a phase
  transition between a low temperature / low force compact (folded)
  structure and a high temperature / high force molten (unfolded)
  structure. The influence of $\Rc$ on phase diagrams, critical
  exponents, melting, and force extension curves is derived
  analytically.  For vanishing pulling force, only for the limited
  range of loop exponents $2 < \Rc \lesssim 2.479 $ a melting
  transition is possible; for $\Rc\leq2$ the chain is always in the
  folded phase and for $2.479 \lesssim \Rc$ always in the unfolded
  phase.  A force induced melting transition with singular behavior is
  possible for all loop exponents $ \Rc < 2.479 $ and can be observed
  experimentally by single molecule force spectroscopy.  These
  findings have implications for the hybridization or denaturation of
  double stranded nucleic acids.  The Poland-Scheraga model for
  nucleic acid duplex melting does not allow base pairing between
  nucleotides on the same strand in denatured regions of the double
  strand. If the sequence allows these intra-strand base pairs, we
  show that for a realistic loop exponent $\Rc\approx2.1$ pronounced
  secondary structures appear inside the single strands. This leads to
  a lower melting temperature of the duplex than predicted by the
  Poland-Scheraga model.  Further, these secondary structures
  renormalize the effective loop exponent~$\RcDNA$, which
  characterizes the weight of a denatured region of the double strand,
  and thus affect universal aspects of the duplex melting transition.
}

\date{\today}

\maketitle


\keywords{RNA, DNA, nucleic acids, denaturation, pulling, melting,
  loop entropy, phase transition}


\section{Introduction}
\label{sec:introduction}

Ribonucleic acids continue to stay in the focus of experimentalists
and theorists~\cite{Gesteland2005}. The advance of single molecule
techniques~\cite{Smith1996,Liphardt2002,Rief1999,Maier2000,Bockelmann1997,Mossa2009}
nowadays allows to study single chains of nucleic acids under tension
and varying solution conditions and thereby yields unprecedented
insights into the behavior and folding properties of these essential
molecules.  Theory on RNA folding vastly relies on the idea of
hierarchical folding proposed by
Tinoco \emph{et al.}~\cite{Tinoco1971,Tinoco1999}, %
stating that given a sequence (the primary structure), the secondary
structure (\ie the list of all base pairs) forms independently of the
tertiary structure (the overall three-dimensional arrangement of all
atoms). This is in contrast to the protein folding problem, which does
not feature these well separated energy scales between the different
structural levels and hence is more involved~\cite{Finkelstein2004}.
The idea of hierarchical folding therefore suggests to solely focus on
the secondary structure, \ie the base pairs, and to neglect the
tertiary structure. This constitutes a major simplification and
enables to calculate partition functions exactly and to predict the
secondary structure formed by a given RNA sequence.
De~Gennes~\cite{Gennes1968} %
was the first to calculate the partition function of an ideal
homopolymeric RNA chain by using a propagator formalism and solving
the partition function by means of a singularity analysis of the
generating functions. Due to his real space approach for an ideal
polymer, the loop exponent was fixed at $\Rc=3/2$.  The loop
exponent~$\Rc$ characterizes the logarithmic entropy contribution
$\propto\ln\Rm^{-\Rc}$ of a loop of length $\Rm$.  Ten years later,
Waterman and Smith~\cite{Waterman1978a} %
devised a recursion relation appropriate for the partition function of
folded RNA, which now lies at the heart of most RNA secondary
structure and free energy prediction algorithms currently used.  Since
all results obtained for RNA are also valid for DNA on our relative
primitive level of modeling, we will mostly explicitly refer to RNA in
our paper but note that in principle all our results carry over to
single stranded DNA molecules, as well.  Subsequently, several
theoretical models were developed to study RNA and DNA: those were
focused on melting~\cite{Einert2008,Hofacker1994,McCaskill1990},
stretching~\cite{Einert2010,Montanari2001,Gerland2001,Mueller2002,Hanke2008,Cocco2002a},
unzipping~\cite{Lubensky2002}, translocation~\cite{Bundschuh2005},
salt influence~\cite{Einert2010a,Tan2006,Mamasakhlisov2007},
pseudoknots~\cite{Baiesi2003,Orland2002}, and the influence of the
loop exponent~\cite{Einert2008,Einert2010,Blossey2003,Kafri2000}. In
this context, an interesting question in connection with the melting
of double stranded DNA (dsDNA) arises: Do secondary structure elements
form in the single strands inside denatured dsDNA loops or not?
Formation of such secondary structures in dsDNA loops would mean that
inter-strand base pairing between the two strands --~being responsible
for the assembly of the double helix~-- is in competition with
intra-strand pairing, where bases of the same strand interact. This
question is not only important for the thermal melting of dsDNA but
also for DNA transcription, DNA replication and the force-induced
overstretching transition of DNA~\cite{Alberts2002,Rief1999}.

In this paper, the influence of the loop exponent $\Rc$ on the
behavior of RNA subject to varying temperature and external force is
studied, which goes beyond our previous work where only the
temperature influence was considered~\cite{Einert2008}.  We neglect
sequence effects and consider a long homopolymeric, single stranded
RNA molecule. A closed form expression for the partition function is
derived, which allows to study the thermodynamic behavior in
detail. The phase diagram in the force-temperature plane is
obtained. We find that the existence of a temperature induced phase
transition depends crucially on the value of the loop exponent $\Rc$:
at vanishing force a melting transition is possible only for the
limited range of loop exponents $2 < \Rc < 2.479 $. $\Rc\approx2.1$ is
a typical exponent that characterizes the entropy of loops usually
encountered in RNA structures --~hairpin loops, internal loops,
multi-loops with three or more emerging helices. That means that RNA
molecules are expected to experience a transition between a folded and
an unfolded state.
This is relevant for structure formation in DNA or RNA single strands
and can in principle be tested in double laser trap force clamp
experiments~\cite{Gebhardt2010}.  Our findings also have implications
for the denaturation of double helical nucleotides (\eg dsDNA). Since
intra-strand and inter-strand base pairing compete, secondary
structure formation of the single strands inside denatured regions of
the duplex has to be taken into account in a complete theory of dsDNA
melting.  In the case where intra- and inter-strand base pairing
occurs, the classical Poland-Scheraga mechanism for the melting of a
DNA duplex has to be augmented by the single-strand folding scenario
considered by us, as the Poland-Scheraga theory is only valid in the
case where no intra-strand base pairing is possible.  If the
intra-strand interactions are strong enough to induce folded secondary
structures, we show that the loop exponent governing the entropy of
inter-strand loops is renormalized and takes on an effective universal
value that only depends on whether the inter-strand loops are
symmetric (consisting of two strands of the same length) or
asymmetric. The resulting duplex melting transition is universal and
turns out to be strongly discontinuous in the first, symmetric case,
and on the border between continuous and discontinuous in the second,
asymmetric case.  In the case when intra-strand base pairing is weak
and a single strand is in the unfolded phase, the situation is
different and qualitatively similar to the original Poland-Scheraga
results, yet with a lower melting temperature.  All these effects can
be studied experimentally. We make explicit suggestions for dsDNA
sequences, with which the formation of intra-strand secondary
structures inside inter-strand loops can be selectively inhibited or
favored.

\section{Derivation of partition function}
\label{sec:deriv-part-funct}

\subsection{Model}
\label{sec:model}

\begin{figure}
  \centering
  \includegraphics{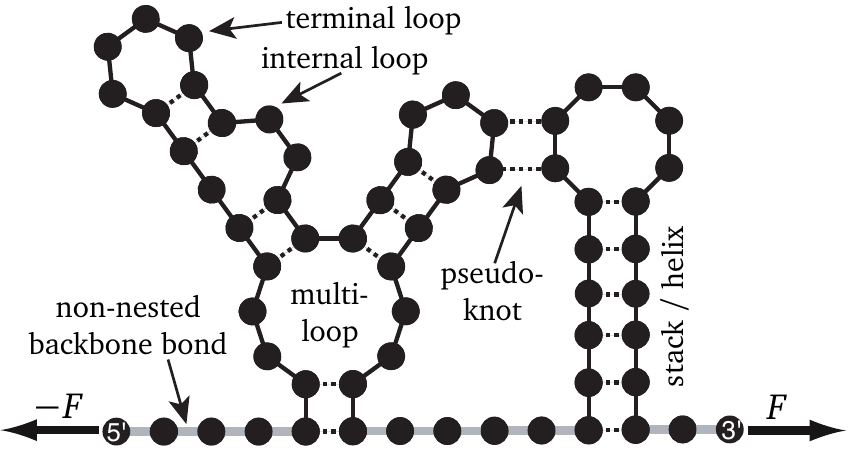}
  \caption{Schematic representation of the secondary structure of an
    RNA molecule. Dots represent one base, \ie cytosine, guanine,
    adenine, or uracil. Solid lines denote the sugar-phosphate
    backbone bonds, broken lines base pairs, and thick gray lines the
    non-nested backbone bonds, which are counted by the variable
    $\RM$, here $\RM=11$. The thick arrows to either side illustrate
    the force~$\RF$ applied to the 5'- and 3'-end. }
  \label{fig:1}
\end{figure}

\begin{figure}
  \centering
  \includegraphics{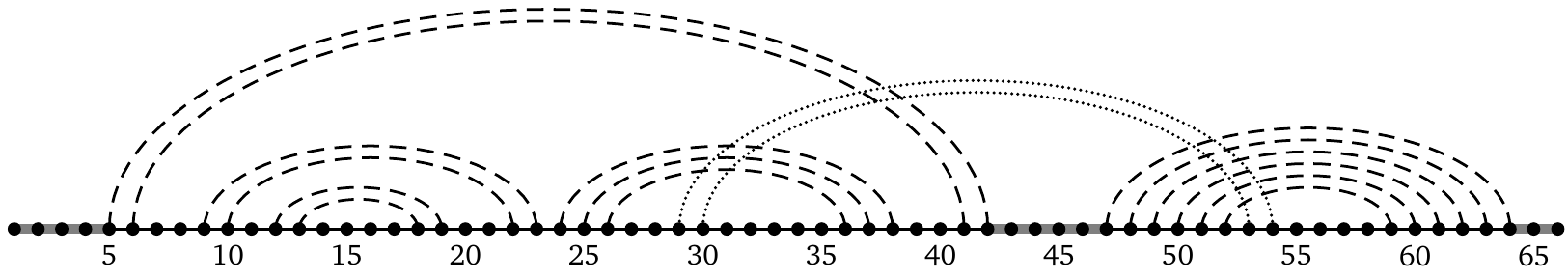}
  \caption{The arc diagram is a representation of the secondary
    structure depicted in \fig~\ref{fig:1}. A dot represents one
    base. Solid lines denote the backbone bonds and thick gray lines
    the non-nested backbone bonds. Dashed arcs denote hydrogen bonds
    between two bases. A pseudoknot (dotted arc) is recognized here as
    crossing arcs. If no pseudoknots are present the structure is
    hierarchical, meaning that substructures are either nested or
    juxtaposed.}
  \label{fig:2}
\end{figure}

Single stranded RNA is modeled as a one-dimensional chain. The basic
units are nucleotides with the four bases (cytosine~(C), adenine~(A),
guanine~(G), uracil~(U)), which are enumerated by an index $i =
1,\ldots,\RN$. A nucleotide can establish a base pair~(bp) with
another nucleotide \via hydrogen bonding, leading to helices and loops
as the structural building blocks of an RNA secondary structure, see
\fig~\ref{fig:1}. In agreement with previous treatments, a valid
secondary structure is a list of all base pairs with the constraint
that a base can be part of at most one pair. In addition, pseudoknots
are not allowed, that means that for any two base pairs $\Rbp{i}{j}$
and $\Rbp{k}{l}$ with $i<j$, $k<l$, and $i<k$ we have either $i<k<l<j$
or $i<j<k<l$~\cite{Orland2002}.  This imposes a hierarchical order on
the base pairs, meaning that two base pairs are either nested and part
of the same substructure or are independent and part of different
substructures, \fig~\ref{fig:2}.  Helix stacking --~the interaction of
two helices emerging from the same loop~-- and base triples as well as
the overall three-dimensional structure are not considered.  Base
pairs are stabilized by two different interactions. First, by hydrogen
bonds between complementary bases and, second, by the stacking
interaction between neighboring base pairs, which are accounted for by
the sequence independent parameters $\RghHB$ and $\Rghstack$,
respectively. A helix with $h$ base pairs consequently has the free
energy $h (\RghHB + \Rghstack) - \Rghstack + \Rghi$, where $\Rghi$ is
a helix initiation free energy.  Therefore, the hydrogen bonding and
the stacking interaction can be combined to yield the binding energy
per base pair $\Reps = -(\RghHB + \Rghstack)$, which we define to be
positive. $\Reps$ can be measured experimentally by duplex
hybridization~\cite{Xia1998} and contains the binding free energies as
well as the extensive part of configurational polymer entropy.
Further, the stacking interaction appears as an additional
contribution to the helix initiation free energy.  We define $\RGhinit
= \Rghi - \Rghstack$ and describe the binding free energy by a single,
sequence independent parameter~\cite{Einert2008}
\begin{equation}
  \label{eq:1}
  \Rw=\exp(\Reps/(\kBT))
  \eqspace.
\end{equation}
$\Rw$ is the statistical weight of a bound base pair, $\RT$ is the
absolute temperature, and $\kB$ the Boltzmann constant.  Therefore
this is a model for homopolymeric RNA, which can be realized
experimentally with synthetic alternating sequences
$[\mathrm{AU}]_{\RN/2}$ or $[\mathrm{GC}]_{\RN/2}$. It has also been
argued that this homopolymeric model describes random RNA above the
glass transition~\cite{Bundschuh2002a}.

The non-extensive contribution of the free energy of a loop is given
by
\begin{equation}
  \label{eq:2}
  \RGlentropy = - \kBT \ln \Rm^{-\Rc}
\end{equation}
and describes the entropy difference between an unconstrained polymer
and a looped polymer. The loop exponent~$\Rc$ is
$\Rc_{\mathrm{ideal}}=3/2$ for an ideal polymer and
$\Rc_{\mathrm{SAW}}=d\nu \simeq 1.76$ for an isolated self avoiding
loop with $\nu \simeq 0.588$ in $d=3$ dimensions~\cite{Gennes1979}.
However, helices, which emerge from the loop, increase $\Rc$ even
further.  In the asymptotic limit of long helical sections,
renormalization group predicts $\Rc_l = d\nu + \sigma_l-l\sigma_3$ for
a loop with $l$ emerging helices~\cite{Duplantier1986,Kafri2000},
where $\sigma_l=\epsilon l(2-l) /16 +\epsilon^2 l (l-2)(8l-21)/512
+\mathcal O (\epsilon^3)$ in an $\epsilon=4-d$ expansion.  One obtains
$\Rc_1=2.06$ for terminal, $\Rc_2=2.14$ for internal loops and
$\Rc_4=2.16$ for a loop with four emerging helices.  For larger $l$
the $\epsilon$ expansion prediction for $\Rc_l$ becomes unreliable.
One sees that the variation of $\Rc$ over different loop topologies
that appear in the native structures of RNA is quite small, which
justifies our usage of the same exponent $\Rc$ for loops of all
topologies that occur in a given RNA secondary structure.

\subsection{Canonical partition function}
\label{sec:canon-part-funct}
\begin{figure}
  \centering
  \includegraphics{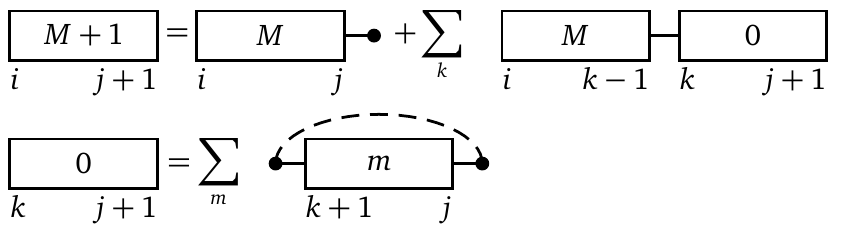}
  \caption{Illustration of the recursion scheme for the canonical
    partition function in \eq~\eqref{eq:3}. Boxes denote partition
    functions of substrands (the range is given by the
    subindices). The numbers inside a box give the number of
    non-nested backbone bonds. To calculate the partition function of
    a strand ranging from $i$ through $j+1$, consider the partition
    function of a strand ranging from $i$ through $j$ and add base
    number $j+1$, which may (right term in first row) or may not (left
    term in first row) establish a base pair with base number $k$. In
    the second row $ \RQH_{k,j+1}^{0}$ is calculated by closing
    structures with $\Rm$ non-nested bonds with a hydrogen bond
    (dashed line). For homopolymeric RNA, the sequence dependence
    drops out and only the lengths of the substrands, $\RN = j-i$, $n
    = k-1-i$, $\RN-n = j+1-k$, need to be considered, see
    \eq~\eqref{eq:3}.}
  \label{fig:3}
\end{figure}
As we neglect pseudoknots, only hierarchical structures are present,
which allows to write down a recursion relation for the partition
function. Further, as we consider homopolymers and omit sequence
effects by using a constant base pairing weight~$\Rw$, the system is
translationally invariant. Hence, the canonical partition function
$\RQH_{i,j}^{\RM}$ of a strand ranging from base $i$ at the 5'-end
through $j$ at the 3'-end depends only on the total number of segments
$\RN = j-i$ and on the number of non-nested backbone bonds $\RM$. A
non-nested bond is defined as a backbone bond, which is neither part
of a helix nor part of a loop. It is outside all secondary structure
elements and therefore contributes to the end-to-end extension, which
couples to an external stretching force and which can be observed for
example in force spectroscopy
experiments~\cite{Einert2008,Bundschuh2005,Mueller2003}, see
\figs~\ref{fig:1} and~\ref{fig:2}.  The recursion relations for
$\RQH_{\RN}^{\RM}$ can be written as
\begin{subequations}\label{eq:3}
  \begin{equation}
    \label{eq:3a}
    \RQH_{\RN+1}^{\RM+1} = \frac{\Rvf{\RM+1}}{\Rvf{\RM}} \left[ \RQH_{\RN}^{\RM} +
      \Rw\sum_{n=\RM}^{\RN-1}\RQH_{n}^{\RM} \RQH_{N-n}^{0}\right]
  \end{equation}
  and
  \begin{equation}
    \label{eq:3b}
    \RQH_{\RN-n}^{0}= \sum_{\Rm=-1}^{\RN-n-2}\frac{\RQH_{\RN-n-2}^{\Rm}}{\Rvf{m}}
    \frac{\exp\bigl(-\RGhinit\HeavisideTheta{\Rm-2}/(\kBT)\bigr)}{(\Rm+2)^\Rc}
    \eqspace
  \end{equation}
\end{subequations}
and is illustrated in \fig~\ref{fig:3}. The Heaviside step function is
$\HeavisideTheta{\Rm} = 0$ if $\Rm \leq 0$ and $\HeavisideTheta{\Rm} =
1$ if $\Rm >0$.  \Eq~\eqref{eq:3a} describes the elongation of an RNA
structure by either adding an unpaired base (first term) or by adding
an arbitrary substrand $ \RQH_{\RN-n}^{0}$ that is terminated by a
helix.  \Eq~\eqref{eq:3b} constructs $ \RQH_{\RN-n}^{0}$ by closing
structures with $\Rm$ non-nested bonds, summed up in $
\RQH_{\RN-n-2}^{\Rm}$, by a base pair.  $\Rvf{\RM}$ denotes the number
of configurations of a free chain with $\RM$ links and can be
completely eliminated from the recursion relation by introducing the
rescaled partition function
$\RQHt_{\RN}^{\RM}=\RQH_{\RN}^{\RM}/\Rvf{\RM}$. We set $\RGhinit = 0$
for computational simplicity and combine \eqs~\eqref{eq:3a}
and~\eqref{eq:3b}, which leads to the final recursion relation
\begin{equation}
  \label{eq:4}
  \RQHt_{\RN+1}^{\RM+1} = \RQHt_{\RN}^{\RM} +
  \Rw\sum_{n=\RM}^{\RN-1}
  \sum_{\Rm=-1}^{\RN-n-2}\frac{\RQHt_{n}^{\RM}\RQHt_{\RN-n-2}^{\Rm}}{(\Rm+2)^\Rc}
  \eqspace,
\end{equation}
with the boundary conditions $\RQHt_{-1}^{-1} = 1$, $\RQHt_{\RM}^{\RN}
= 0$ for $\RM>\RN$, $\RN<0$, or $\RM<0$. The thermodynamic limit of an
infinitely long RNA chain is described by the canonical Gibbs
ensemble, which is characterized by a fixed number of segments $\RN$,
but a fluctuating number of non-nested backbone bonds~$\RM$. Therefore
we introduce the unrestricted partition function
\begin{equation}
  \label{eq:5}
  \RQG_{\RN}(\Rs) = \sum_{\RM = 0}^\infty \Rs^\RM \RQHt_{\RN}^{\RM}
  \eqspace,
\end{equation}
which contains the influence of an external force~$\RF$ \via the
statistical weight~$\Rs$ of a non-nested backbone bond.  For RNA with
no force applied to the ends, one has $\Rs = 1$. We model the RNA
backbone elasticity by the freely jointed chain (FJC) model, where the
weight of a non-nested backbone bond subject to an external force is
given by
\begin{equation}
  \Rs
  = \frac{1}{4\pi}\int_0^{2\pi} \drm \phi \int_0^\pi \drm\theta\productspace \e^{-\beta \RF\RKuhnss \cos\theta}
  = \frac{\sinh(\beta\RF\RKuhnss)}{\beta\RF\RKuhnss}\label{eq:6}
  \eqspace;
\end{equation}
here, we introduced the inverse thermal energy $\beta = (\kBT)^{-1}$
and the Kuhn length~$\RKuhnss$.

\subsection{Grand canonical partition function}
\label{sec:grand-canon-part}

For studying the phase transition and the critical behavior, it is
useful to introduce the generating function or grand canonical
partition function
\begin{equation}
  \label{eq:7}
  \RZG(\Rz,\Rs) = \sum_{\RN = 0}^\infty \Rz^\RN \RQG_{\RN}(\Rs)
  = \sum_{\RN = 0}^\infty \sum_{\RM = 0}^\infty \Rz^\RN \Rs^\RM \RQHt_{\RN}^{\RM}
  \eqspace,
\end{equation}
where $\Rz = \exp(\Rmu/(\kBT))$ is the fugacity.
\begin{figure}
  \centering
  \includegraphics{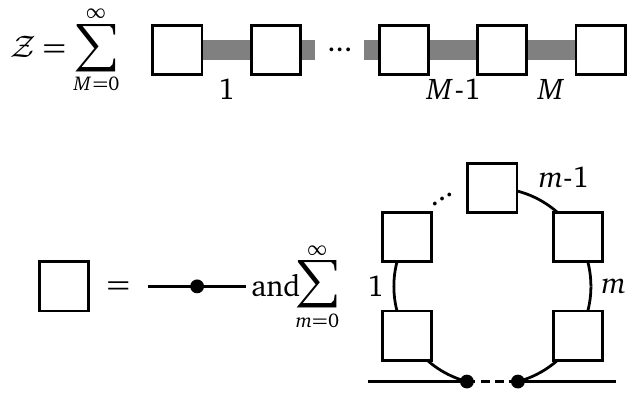}
  \caption{Structure of the grand canonical partition function
    $\RZG(\Rz,\Rs)$ according to \eqs~\eqref{eq:7}
    and~\eqref{eq:11}. The grand canonical partition function of the
    Gibbs ensemble is a sum over all numbers of non-nested backbone
    bonds (thick gray lines) with statistical weight $\Rs\Rz$. Between
    two adjacent backbone bonds can be either a single nucleotide
    (dot), with statistical weight $1$, or a structure with weight
    $\RZGb$, whose first and last base are paired. The white squares
    serve as wildcards for either possibility and have the statistical
    weight $\Rk(\Rw,\Rz)$. Thin black lines depict backbone bonds that
    are part of a helix or loop and have statistical weight $\Rz$. }
  \label{fig:4}
\end{figure}
Performing the weighted double sum $\sum_{\RN = -1}^\infty \sum_{\RM =
  -1}^\infty \Rz^\RN \Rs^\RM$ on both sides of \eq~\eqref{eq:4} yields
\begin{equation}
  \label{eq:8}
  (\Rs\Rz)^{-1}\RZG = (\Rs\Rz)^{-1} + \RZG + \bigl((\Rs\Rz)^{-1} + \RZG\bigr)(\Rk-1)
  \eqspace,
\end{equation}
which can be solved and one obtains the generating function
\begin{equation}
  \label{eq:9}
  \RZG(\Rz,\Rs) = \frac{\Rk(\Rw,\Rz)}{1-\Rs\Rz\Rk(\Rw,\Rz)}
  \eqspace.
\end{equation}
Here we have defined $\Rk(\Rw,\Rz) = 1+\RZGb(\Rw,\Rz)$ as the grand
canonical partition function of RNA structures with zero non-nested
backbone bonds, \ie structures which consist of just one nucleotide or
structures where the terminal bases are paired,
\begin{equation}
  \label{eq:10}
  \Rk(\Rw,\Rz) = 1 + \RZGb(\Rw,\Rz)
  = 1+ \Rw\Rz^2\sum_{\RN=-1}^{\infty}\sum_{\RM=-1}^{\RN} \Rz^\RN
  \frac{\RQHt_{\RN}^{\RM}}{(\RM+2)^{-\Rc}}
  \eqspace.
\end{equation}
\Eq~\eqref{eq:9} has an instructive interpretation, which becomes
clear by expanding the fraction in a geometric series
\begin{equation}
  \label{eq:11}
  \RZG(\Rz,\Rs) = \sum_{\RM=0}^\infty \Rs^\RM\productspace\Rz^\RM\Rk^{\RM+1}
  =  \sum_{\RM=0}^\infty (1+\RZGb)\cdot\bigl(\Rs\Rz (1+\RZGb)\bigr)^{\RM}
  \eqspace,
\end{equation}
where $\Rs\Rz$ is the statistical weight for a backbone segment which
is not nested.  Between two adjacent segments we have the possibility
to put either a single nucleotide (with statistical weight $1$) or a
structure whose first and last bases are paired (with statistical
weight $\RZGb$). See \fig~\ref{fig:4} for an illustration.

\begin{figure}
  \centering
  \includegraphics{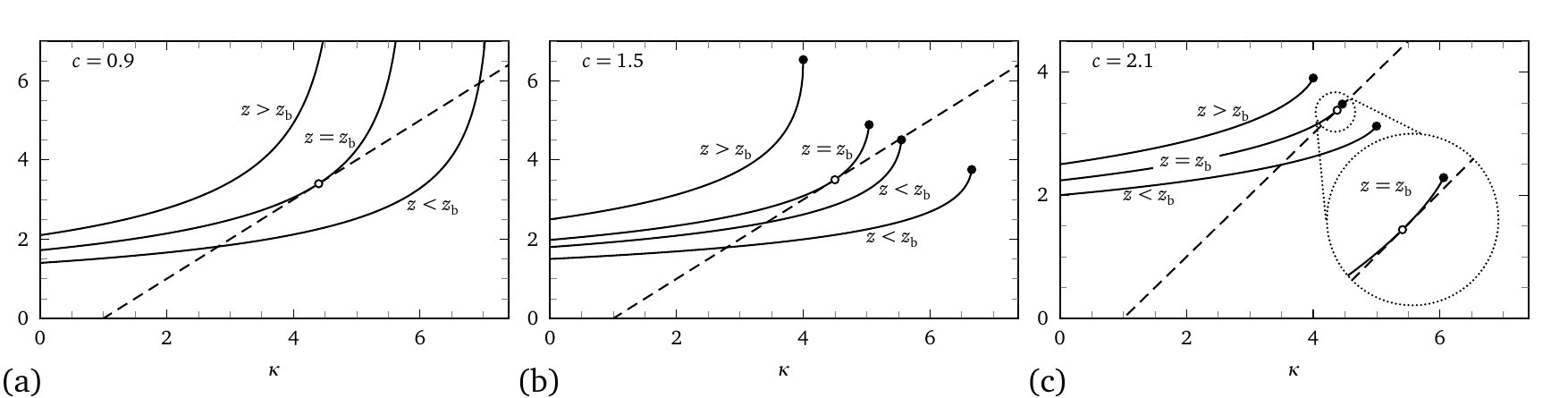}%
  \subfloat{\label{fig:5a}}%
  \subfloat{\label{fig:5b}}%
  \subfloat{\label{fig:5c}}%
  \caption{ Graphical solution of the equations, which determine
    $\Rk(\Rw,\Rz)$, $\Rzb$, $\Rzp$. The functions $\Rk-1$ (dashed
    line) and $\Rh(\Rk,\Rz)$ (solid lines), \eq~\eqref{eq:13}, are
    plotted for $\Rw=10$ and different values of the fugacity $\Rz$
    and the loop exponent~$\Rc$, \subref{fig:5}a~$\Rc = 0.9$
    \subref{fig:5}b~$\Rc = 3/2$ \subref{fig:5}c~$\Rc = 2.1$. Points at
    which both curves intersect are solutions of \eq~\eqref{eq:12} and
    determine $\Rk(\Rw,\Rz)$. In case of two positive solutions the
    smaller yields the correct solution as $\Rk(\Rw,\Rz)$ is to be a
    continuous, monotonically increasing function of $\Rz$ with
    $\Rk(\Rw,0) = 1$, \eq~\eqref{eq:10}.  Points at which both curves
    are adjacent to each other (open circles) determine the branch
    point~$\Rzb$. Points at which $\Rz\Rk=1$ (filled circles)
    determine the position of the pole~$\Rzp$ in the absence of force,
    $\Rs=1$. }
  \label{fig:5}
\end{figure}

In order to determine the function $\Rk(\Rw,\Rz)$, we compare the
coefficients of the power series in $\Rs$ in \eqs~\eqref{eq:7}
and~\eqref{eq:11} and obtain $\Rz^\RM\Rk^{\RM+1} =
\sum_{\RN=\RM}^\infty \Rz^\RN \RQHt_{\RN}^{\RM}$. The lower summation
index is due to exchanging the summations in \eq~\eqref{eq:7}, bearing
in mind that $ \RQHt_{\RN}^{\RM} = 0$ for $\RM > \RN$.  This identity
can be inserted into \eq~\eqref{eq:10} and yields the equation
\begin{equation}
  \label{eq:12}
  \Rk(\Rw,\Rz) -1 = \frac{\Rw}{\Rk(\Rw,\Rz)}\Li{\Rc}{\Rz\Rk(\Rw,\Rz)}
  \eqspace,
\end{equation}
which determines $\Rk(\Rw,\Rz)$. $\Li{\Rc}{\Rz\Rk}=\sum_{\Rm=1}^\infty
\Rz^\Rm \Rk^\Rm/m^{\Rc}$, for $\Rz\Rk\leq1$, is the
polylogarithm~\cite{Erdelyi1953}.  We introduce
\begin{equation}
  \label{eq:13}
  \Rh(\Rk,\Rz) = \frac{\Rw}{\Rk}\Li{\Rc}{\Rz\Rk}
  \eqspace
\end{equation}
and rewrite \eq~\eqref{eq:12} as $ \Rk -1 = \Rh(\Rk,\Rz) $.
\Eq~\eqref{eq:12} has at most two positive and real solutions as can
be seen from \fig~\ref{fig:5}, where we plot the two sides of
\eq~\eqref{eq:12}. $\Rk(\Rw,\Rz)$ is a continuous, monotonically
increasing function of $\Rz$ with $\Rk(\Rw,0) = 1$ as follows from
\eq~\eqref{eq:10}. Therefore, only the smallest positive root of
\eq~\eqref{eq:12} yields the correct $\Rk(\Rw,\Rz)$.  For $\Rz
\rightarrow 0$ there is always a positive and real solution for
$\Rk(\Rw,\Rz)$.  Increasing $\Rz$ increases $\Rk(\Rw,\Rz)$ until
eventually, at $\Rz = \Rzb$, the real solution for $\Rk(\Rw,\Rz)$
vanishes. Thus, $\Rk(\Rw,\Rz)$ has a branch point at $\Rz = \Rzb$.
Depending on the value of the loop exponent~$\Rc$, the polylogarithm
$\Li{\Rc}{\Rz\Rk}$ and $\Rh(\Rk,\Rz)$ \subref{fig:5}a~are divergent
for $\Rc\leq1$, \subref{fig:5}b~are finite, but feature a diverging
slope for $1<\Rc\leq2$, or \subref{fig:5}c~have a finite value and
derivative for $2<\Rc$ at $\Rz\Rk=1$, see \fig~\ref{fig:5}. This will
become important later, when the existence of phase transitions is
studied.
 
\subsection{Back-transform to canonical ensemble}
\label{sec:backtr-canon-ensembl}

Since the thermodynamic limit $\RN\rightarrow \infty$ is defined in
the canonical Gibbs ensemble, we now demonstrate how to obtain
$\RQG_{\RN}(\Rs)$, \eq~\eqref{eq:5}, from $\RZG(\Rz,\Rs)$,
\eq~\eqref{eq:9}.  For large systems, $\RN\gg1$, the canonical
partition function is given by the dominant singularity~$\Rzd(\Rs)$ of
$\RZG(\Rz,\Rs)$, which is defined as the singularity which is nearest
to the origin in the complex
$\Rz$-plane~\cite{Flajolet1990,Einert2010}. In particular if
$\RZG(\Rz, \Rs)\sim K(\Rs)\bigl(\Rzd(\Rs)-\Rz\bigr)^{-\alpha}$ with
$K(\Rs)$ independent of $\Rz$, we obtain
\begin{equation}
  \label{eq:14}
  \RQG_{\RN}(\Rs) \sim  \Rzd^{-\RN}(\Rs)\RN^{\alpha-1}\cdot K(\Rs)  \Rzd^{-\alpha}(\Rs)/\Gamma(\alpha)
  \eqspace,
\end{equation}
where $\Gamma(\alpha)$ is the gamma
function~\cite{Abramowitz2002}. Therefore, the Gibbs free energy reads
to leading orders in $\RN$
\begin{equation}
  \label{eq:15}
  \RGG/(\kBT) = - \ln \RQG_{\RN} \sim \RN \ln \Rzd(\Rs) + (1-\alpha)\ln\RN
  \eqspace.
\end{equation}

\label{sec:type-singularities}
In fact, $\RZG(\Rz,\Rs)$ features two relevant singularities. First,
the branch point~$\Rzb(\Rw)$ of $\Rk(\Rw,\Rz)$, which is independent
of $\Rs$, and second a simple pole~$\Rzp(\Rw,\Rs)$, where the
denominator of $\RZG(\Rz,\Rs)$ vanishes, see
\eq~\eqref{eq:9}. Depending on which singularity has the smallest
modulus, the molecule can be in different phases. In the following
sections it will turn out that the low temperature, compact or folded
phase is associated with $\Rzb$, whereas the high temperature,
extended or unfolded phase is characterized by $\Rzp$.

Let us consider the branch point first.  It can be seen from
\fig~\ref{fig:5}, that for $\Rz<\Rzb$ at least one real solution of
\eq~\eqref{eq:12} exists, where the smaller solution determines
$\Rk(\Rw,\Rz)$. Right at $\Rz=\Rzb$ the two solutions merge and the
slope of $\Rh$ is $\Rh'(\Rk, \Rzb) = \partial
\Rh(\Rk,\Rzb)/\partial\Rk = 1$ at the tangent point. This yields the
equation for the position of the branch point singularity $\Rzb(\Rw)$,
which is a function of $\Rw$ only,
\begin{equation}
  \label{eq:16}
  \Rk(\Rw,\Rzb)^2=\Rw\Li{\Rc-1}{\Rzb\Rk(\Rw,\Rzb)} - \Rw\Li{\Rc}{\Rzb\Rk(\Rw,\Rzb)}
  \eqspace.
\end{equation}
The behavior of $\Rk(\Rw,\Rz)$ in the vicinity of the branch point can
be obtained by expanding \eq~\eqref{eq:12} for $\Rz\rightarrow\Rzb$
and $\Rzb\Rk(\Rw,\Rzb)<1$
\begin{equation}
  \label{eq:17}
  \Rk(\Rw,\Rz) \sim \Rkb - \Bigl({\frac{\Rzb-\Rz}{\Rzb}}\Bigr)^{1/2}
  \RKb {(1-\Rs\Rzb\Rkb)^2}
  \eqspace,
\end{equation}
where we used the short notation $\Rkb = \Rk(\Rw,\Rzb)$ and defined
\begin{equation}
  \label{eq:58}
  \RKb = \Bigl({\frac{2\Rw\Li{\Rc-1}{\Rzb\Rkb}}{ \Rw\Li{\Rc-2}{\Rzb\Rkb} -
      \Rw\Li{\Rc-1}{\Rzb\Rkb} - 2\Rkb^2 }}\Bigr)^{1/2} \frac{\Rkb}{(1-\Rs\Rzb\Rkb)^2}
\end{equation}
Due to the exponent $1/2$ in the above equation, the function
$\Rk(\Rw,\Rz)$ exhibits a first order branch point at $\Rz = \Rzb$ and
the grand canonical partition function, \eq~\eqref{eq:9}, scales as
\begin{equation}
  \label{eq:18}
  \RZG(\Rz,\Rs) \sim \frac{\Rkb}{1-\Rs\Rzb\Rkb} -\Bigl({\frac{\Rzb-\Rz}{\Rzb}}\Bigr)^{1/2} \RKb
  \eqspace.
\end{equation}
Together with \eq~\eqref{eq:14}, we obtain the following scaling for
the canonical partition function
\begin{equation}
  \label{eq:19}
  \RQG_{\RN}(\Rs)\sim \Rzb^{-\RN} \RN^{-3/2}  \RKb / \sqrt{4\pi}
  \eqspace,
\end{equation}
which leads to a logarithmic $\RN$-contribution with universal
prefactor $3/2$ to the free energy $\RGG = -\kBT\ln\RQG_{\RN}$, in
accord with the findings of
de~Gennes~\cite{Gennes1968}. %
It will turn out that \eq~\eqref{eq:19} describes the low temperature
or folded phase of the system.

Now let us consider the pole singularity $\Rzp$ of the grand canonical
partition function. $\Rzp(\Rw,\Rs)$ is a function of $\Rw$ and $\Rs$
and is given by the zero of the denominator of $\RZG(\Rz,\Rs)$ in
\eq~\eqref{eq:9},
\begin{equation}
  \label{eq:20}
  \Rs\Rzp\Rk(\Rw,\Rzp) = 1
  \eqspace.
\end{equation}
The position of the pole can be evaluated in a closed form expression
by plugging \eq~\eqref{eq:20} into \eq~\eqref{eq:12} and solving the
resulting quadratic equation for $\Rz$. One obtains
\begin{subequations}
  \label{eq:21}
  \begin{gather}
    \label{eq:17a}
    \Rzp(\Rw,\Rs) = \frac{2}{\Rs}\left(1+\sqrt{1+4\Rw\Li{c}{1/\Rs}}\right)^{-1}\eqspace,\\
    \label{eq:17b}
    \Rk(\Rw,\Rzp) =
    \frac{1}{2}\left(1+\sqrt{1+4\Rw\Li{c}{1/\Rs}}\right) \eqspace.
  \end{gather}
\end{subequations}
The behavior of $\Rk(\Rw,\Rz)$ in the vicinity of the pole can be
obtained by expanding \eq~\eqref{eq:12} for $\Rz\rightarrow\Rzp$,
$\Rzp\Rk(\Rw,\Rzp)=1/\Rs<1$, and $\Rc>2$
\begin{equation}
  \label{eq:22}
  \Rk(\Rw, \Rz) \sim \Rkp - \Rkp\frac{\Rzp-\Rz}{\Rzp}\timesspace
  \frac{\Rw\Li{\Rc-1}{1/s}}{\RKp (2\Rkp - 1)}
  \eqspace,
\end{equation}
where we used the short notation $\Rkp = \Rk(\Rw,\Rzp)$ and introduced
\begin{equation}
  \label{eq:59}
  \RKp = \frac {2\Rkp^2 - \Rkp - \Rw\Li{\Rc-1}{1/s}}{2\Rkp -1 }
  \eqspace.
\end{equation}
Therefore, the grand canonical partition function scales as
\begin{equation}
  \label{eq:23}
  \RZG(\Rz,\Rs) \sim \Bigl(\frac{\Rzp-\Rz}{\Rzp}\Bigr)^{-1} \RKp
  \eqspace
\end{equation}
and together with \eq~\eqref{eq:14} we obtain the scaling of the
canonical partition function
\begin{equation}
  \label{eq:24}
  \RQG_{\RN}(\Rs) \sim \Rzp^{-\RN} \RKp
  \eqspace.
\end{equation}
Later we will see, that \eq~\eqref{eq:24} describes the denatured high
temperature phase of the system. In contrast to the branch point
phase, \eq~\eqref{eq:19}, no logarithmic contribution to the free
energy is present.

\section{Critical behavior}
\label{sec:critical-behavior}

The two singularities $\Rzb(\Rw)$ and $\Rzp(\Rw,\Rs)$ are smooth
functions of external variables such as temperature~$\RT$ or
force~$\RF$, which enter \via the weight of a base pair $\Rw$,
\eq~\eqref{eq:1}, and the weight of a non-nested backbone bond $\Rs$,
\eq~\eqref{eq:6}.  As the system is described by the singularity,
which is closest to the origin, a phase transition associated with a
true singularity in the free energy, \eq~\eqref{eq:15}, is possible if
these two singularities cross. For that purpose let us shortly review
the three constitutive equations \eqs~\eqref{eq:12},\eqref{eq:16},
and~\eqref{eq:20}.
As observed earlier, the smallest positive root of \eq~\eqref{eq:12}
yields the function $\Rk(\Rw,\Rz)$.  The simultaneous solution of
\eqs~\eqref{eq:12} and~\eqref{eq:16} yields the branch point
$\Rzb(\Rw)$, whereas the simultaneous solution of \eqs~\eqref{eq:12}
and~\eqref{eq:20} yields the pole $\Rzp(\Rw,\Rs)$, which can be
expressed in a closed form, see \eq~\eqref{eq:21}.
\begin{figure}
  \centering
  \includegraphics{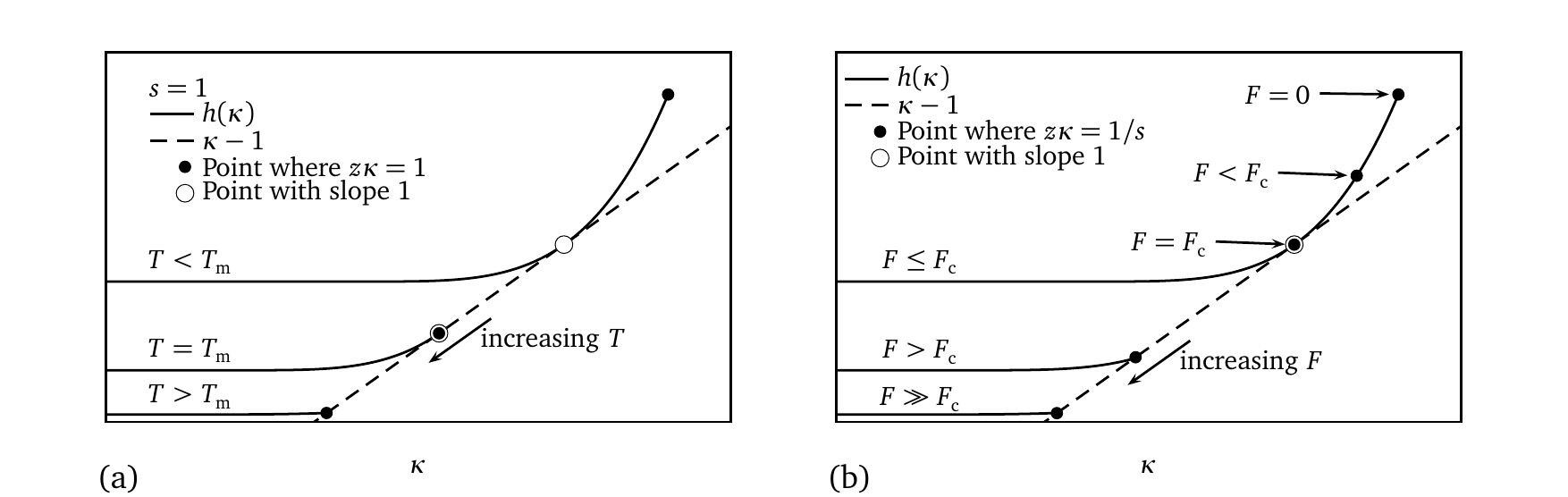}
  \subfloat{\label{fig:6a}} \subfloat{\label{fig:6b}}
  \caption{Illustration of the graphical solution of
    \eqs~\eqref{eq:12},\eqref{eq:16}, and~\eqref{eq:20} at the phase
    transition. The solid line sketches $\Rh(\Rk)$, \eq~\eqref{eq:13},
    the dashed line the function $\Rk-1$. The open circles denote
    points where the conditions for the branch point are met: Curves
    have common points, \eq~\eqref{eq:12}, and curves are tangent to
    each other in these points, \eq~\eqref{eq:16}. The black filled
    circles denote points where $\Rs\Rz\Rk =1$, \eq~\eqref{eq:20}. If
    the black filled circle lies on the dashed curve the conditions
    for a pole are met, \eqs~\eqref{eq:12} and~\eqref{eq:20}. For a
    given temperature $\RT$ and force $\RF$ the fugacity $\Rz$ is
    increased from $\Rz=0$ to the value where either
    \eqs~\eqref{eq:12} and~\eqref{eq:16} hold (open circle on dashed
    line, folded phase) or \eqs~\eqref{eq:12} and~\eqref{eq:20} hold
    (black filled circle on dashed line, unfolded phase).
    \subref{fig:6}a~Illustration of the thermal phase transition at
    zero force, $\Rs =1$. For low temperatures the branch point is
    dominant. Upon increasing the temperature, \ie decreasing $\Rw$,
    the branch point and the pole approach each other until they
    eventually coincide at the melting temperature $\RT=\RTmelt$ and
    cause a phase transition. For $\RT>\RTmelt$ there is no branch
    point anymore and the pole is dominant.
    \subref{fig:6}b~Illustration of the force induced phase transition
    at a temperature $\RT<\RTmelt$. For small forces the branch point
    is dominant. Upon increasing the force, \ie increasing $\Rs$, the
    point where $\Rz\Rk=1/\Rs$ (black filled circle) moves towards the
    branch point. As the branch point is independent of the force, see
    \eqs~\eqref{eq:12} and~\eqref{eq:16}, no observable depends on
    $\RF$ as long as $\RF<\RFc$. At $\RF = \RFc$ the branch point and
    the pole coincide and a phase transition occurs. For $\RF>\RFc$
    the pole is dominant.  }
  \label{fig:6}
\end{figure}

\subsection{Critical point and existence of a phase transition}
\label{sec:crit-point-exist}

The critical fugacity $\Rzc$ and thus the phase transition is defined
as the point where the branch point and the pole coincide
\begin{equation}
  \Rzc = \Rzb(\Rwc) = \Rzp(\Rwc,\Rsc)\label{eq:26}
  \eqspace,
\end{equation}
which means that all three \eqs~\eqref{eq:12},\eqref{eq:16},
and~\eqref{eq:20} have to hold simultaneously, see
\fig~\ref{fig:6}. Assuming that a pair $(\Rwc,\Rsc)$ exists so that
\eq~\eqref{eq:26} is true, this can be evaluated further by plugging
\eqs~\eqref{eq:21} into \eq~\eqref{eq:16} and we obtain
\begin{equation}
  \label{eq:27}
  \Rwc = \frac{\Li{\Rc-1}{1/\Rsc} - \Li{\Rc}{1/\Rsc}}{(\Li{\Rc-1}{1/\Rsc} - 2
    \Li{\Rc}{1/\Rsc})^2}
  \eqspace.
\end{equation}
This constitutes a closed form relation between $\Rwc$ and $\Rsc$ or,
by employing \eqs~\eqref{eq:1} and~\eqref{eq:6}, the critical
temperature~$\RTc$ and force~$\RFc$. The melting temperature $\RTmelt$
is defined as the critical temperature at zero force.

The order of the branch point exactly at $\RT=\RTmelt$ and zero force,
$\Rs = 1$, is calculated by expanding \eq~\eqref{eq:12} in powers of
$\Rz/\Rzc -1$ and $\Rk(\Rw,\Rz)/\Rkc -1$ while keeping $\Rw=\Rwc$
fixed. For vanishing force $\Rkc = 1/\Rzc$ and we obtain
\begin{equation}
  \label{eq:28}
  \Rk(\Rw,\Rz) \sim \Rkc - \frac{\Rkc}{\RKcT}
  \Bigl(\frac{\Rzc-\Rz}{\Rzc} \Bigr)^{{1}/{(\Rc-1)}}
  \eqspace,
\end{equation}
where we used
\begin{equation}
  \label{eq:63}
  \RKcT =   \Bigl(\frac{\RiemannZeta{\Rc-1}}{\Gamma(1-\Rc)}\Bigr)^{-1/(\Rc-1)}
  \eqspace.
\end{equation}
Thus, the asymptotic behavior of the generating function at
$\RT=\RTmelt$ and zero force is
\begin{equation}
  \label{eq:29}
  \RZG(\Rz,\Rs) \sim \RKcT \Bigl(\frac{\Rzc - \Rz}{\Rzc}\Bigr)^{-1/(\Rc-1)} 
  \eqspace,
\end{equation}
and we obtain the modified scaling of the canonical partition function
right at the melting point temperature
\begin{equation}
  \label{eq:30}
  \RQG_{\RN}({\Rs = 1}) \sim \Rzc^{-\RN} \RN^{(2-\Rc)/(\Rc-1)}  {\RKcT}/{\Gamma({1/(\Rc-1)})}
  \eqspace.
\end{equation}
We see that the loop statistics are crucial and enter \via the loop
exponent $\Rc$, which gives rise to non-universal critical behavior.

For finite force, $\Rs > 1$, the branch point is first order and the
scaling of the generating function at the critical point reads
\begin{equation}
  \label{eq:31}
  \RZG(\Rz,\Rs) \sim \Bigl(\frac{\Rzc-\Rz}{\Rzc}\Bigr)^{-1/2} \frac{\RKcF}{\Rs}
  \eqspace,
\end{equation}
with
\begin{equation}
  \label{eq:64}
  \RKcF =   \Bigl({\frac { \Rwc\Li{\Rc-2}{1/\Rs} -
      \Rwc\Li{\Rc-1}{1/\Rs} - 2\Rkc^2 }{2\Rwc\Li{\Rc-1}{1/\Rs}}}\Bigr)^{1/2}
  \eqspace,
\end{equation}
leading to the canonical partition function
\begin{equation}
  \label{eq:32}
  \RQG_{\RN}({\Rs}) \sim \Rzc^{-\RN}\RN^{-1/2} \RKcF / (\Rs\sqrt{\pi})
  \eqspace,
\end{equation}
with a scaling independent of the loop exponent $\Rc$.  Note that
\eq~\eqref{eq:31} scales as $(\Rzc-\Rz)^{-1/2}$ in contrast to
\eq~\eqref{eq:18}, which leads to the different scaling of
$\RQG_{\RN}({\Rs})$ in \eq~\eqref{eq:32} when compared to
\eq~\eqref{eq:19}.  In the rest of this section we compare thermal and
force induced phase transition and in particular determine the
parameter range in which a phase transition is possible.

\subsubsection{Thermal phase transition}
\label{sec:therm-phase-trans}

First, we consider the thermal phase transition without external
force, \ie for $\Rs = 1$. In this case the polylogarithm reduces to
the Riemann zeta function, $\Li{\Rc}{1} = \RiemannZeta{\Rc}$.  Since
$\Rs = 1$, we find that $\Rzb \leq \Rzp$ as long as $\Rk(\Rw,\Rz)$ has
a real, positive branch point. This is due to the fact that
$\Rzb\Rk(\Rw,\Rzb) \leq 1$ (see \fig~\ref{fig:5}), $\Rzp\Rk(\Rw,\Rzp)
= 1$, and that $\Rz\Rk(\Rw,\Rz)$ is a monotonically increasing
function of $\Rz$, see \eq~\eqref{eq:10}.

\paragraph{\boldmath No thermal phase transition for $\Rc\leq2$}
\label{sec:no-phase-transition}
For $\Rc\leq2$ the function $\Rk(\Rw,\Rz)$ always features a branch
point since $\Rh'(\Rk) \rightarrow \infty$ for
$\Rk\rightarrow1/\Rz$. This ensures that for every $\Rw$ a $\Rzb$ is
found, where $\Rh(\Rk,\Rzb)$ is tangent to $\Rk-1$, see
\figs~\ref{fig:5}a and~\ref{fig:5}b. As the branch point is always
dominant, we find the universal scaling
\begin{equation}
  \label{eq:33}
  \RGG/(\kBT) = \RN \ln \Rzb + 3/2\ln\RN
\end{equation}
for all temperatures and no phase transition is possible. The RNA
chain is always in the folded phase.

\paragraph{\boldmath No thermal phase transition for $\Rc\geq\Rcs\approx2.479$}
For $\Rc>2$ the function $\Rh(\Rk,\Rz)$ and its derivative are finite
for $\Rz\Rk = 1$.  A sufficient condition for a branch point to exist
is that the slope of $\Rh$ is greater than~$1$ for $\Rz\Rk=1$, see
filled circles in \fig~\ref{fig:5}c, and hence
\begin{equation}
  \label{eq:34}
  \Rh'(\Rk(\Rw,\Rzp),\Rzp) =
  \frac{2\Rw(\RiemannZeta{\Rc-1}-2\RiemannZeta{\Rc})}{1+\sqrt{1+4\Rw\RiemannZeta{\Rc}}}\stackrel{!}{>}1
  \eqspace.
\end{equation}
This can be achieved always for large enough~$\Rw$ as long as the
numerator is positive.  On the other hand, for $\Rc \geq \Rcs
\approx2.479$, where $\Rcs$ is the root of
\begin{equation}
  \label{eq:35}
  \RiemannZeta{\Rcs-1}-2\RiemannZeta{\Rcs} = 0
  \eqspace,
\end{equation}
no branch point exists since the numerator in \eq~\eqref{eq:34} is
negative.  That means that for $\Rc > \Rcs$ the pole $\Rzp(\Rw,\Rs)$
is always the dominant singularity of $\RZG(\Rz,\Rs)$ and the molecule
is always in the unfolded state.

\paragraph{\boldmath Thermal phase transition for $2<\Rc<\Rcs$ at $\Rw
  = \Rwc$}
\label{sec:therm-phase-trans-1}
Only for $2<\Rc<\Rcs$ a thermal phase transition is possible. For
$\Rw>\Rwc$, see \eq~\eqref{eq:27}, the molecule is in the folded phase
governed by the branch point singularity~$\Rzb$, which is determined
by \eqs~\eqref{eq:12} and~\eqref{eq:16}.  Decreasing $\Rw$, \ie
increasing the temperature, causes the branch point and the pole to
approach each other. At the critical point~$\Rwc$, \eq~\eqref{eq:27},
both singularities coincide and a phase transition occurs.  For higher
temperatures, $\Rw<\Rwc$, the RNA is unfolded and described by the
pole~$\Rzp$, \eq~\eqref{eq:21}. See \fig~\ref{fig:6}a for an
illustration. It will turn out that the temperature induced phase
transition at zero force is very weak and that, in fact, the order of
the phase transition is $\RPTn$, where $\RPTn$ is the integer with
$(\Rc-2)^{-1} -1<\RPTn < (\Rc-2)^{-1}$.

\subsubsection{Force induced phase transition}
\label{sec:force-induced-phase}

For the force induced phase transition the situation is slightly
different as the position of the pole $\Rzp$ depends on the force,
which enters \via the weight of a non-nested backbone bond~$\Rs$,
\eq~\eqref{eq:21}. In contrast, the branch point $\Rzb$ does not
depend on $\Rs$ and hence it is constant, \eqs~\eqref{eq:12}
and~\eqref{eq:16}. Therefore, the branch point~$\Rzb$ and the critical
point~$\Rzc$ coincide and $\Rzb$ can be determined exactly by the
relation $\Rzb = \Rzc = \Rzp(\Rwc,\Rsc) = \Tconst$.

\paragraph{\boldmath No force induced phase transition if
  $\Rw<\Rwc(\Rs=1)$ or $\Rc\geq\Rcs$}
\label{sec:boldmath-no-force}

If the molecule is already in the unfolded phase, which can be due to
high temperature, $\Rw<\Rwc$, or due to the non-existence of a branch
point, $\Rc\geq\Rcs$, a force induced phase transition is not
possible. In these cases the pole always dominates the system,
regardless of the value of the applied force.

\paragraph{\boldmath Force induced phase transition if
  $\Rw>\Rwc(\Rs=1)$ and $\Rc<\Rcs$}
\label{sec:boldm-force-induc}

A system below the melting temperature, $\Rw>\Rwc(\Rs=1)$, is in the
folded phase at zero force, $\Rs=1$.  For small forces, \ie
$\Rs<\Rsc$, the system is described by the branch point
singularity~$\Rzb$, which is independent of $\Rs$ and hence does not
depend on the force, \eqs~\eqref{eq:12} and~\eqref{eq:16} and
\fig~\ref{fig:6}b. However, as the pole $\Rzp(\Rw,\Rs)$ is a
monotonically decreasing function of $\Rs$, the branch point and the
pole will eventually coincide at $\Rs = \Rsc$ and a phase transition
occurs.  The critical force fugacity~$\Rsc$ is defined as the root of
\eq~\eqref{eq:27} for fixed $\Rw$. For $\Rs>\Rsc$ the pole is the
dominant singularity and governs the system. Note that in contrast to
the thermal phase transition, a force induced phase transition is
possible even for $\Rc\leq2$. It will turn out that the
  force induced phase transition is second order in accordance with
  previous results \cite{Mueller2003,Montanari2001}.

\subsection{Global phase diagrams}
\label{sec:phase-diagram}
\begin{figure}
  \centering
  \includegraphics{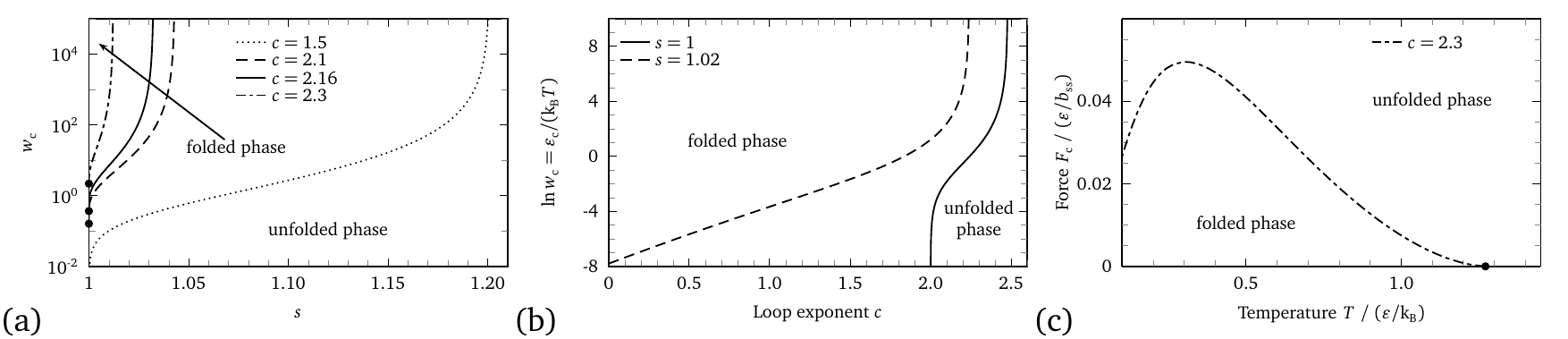}%
  \subfloat{\label{fig:7a}}%
  \subfloat{\label{fig:7b}}%
  \subfloat{\label{fig:7c}}%
 \caption{\subref{fig:7}a~Phase diagram of homopolymeric RNA in the $\Rw$-$\Rs$ plane for
    different values of the loop exponent $\Rc=1.5,\,2.1,\,2.16,\,2.3$ featuring an unfolded
    phase (bottom right) and a folded phase (top left). For $\Rc = \Rcs \approx 2.479$, the
    phase boundary approaches $\Rs = 1$ and the melting point $\Rwc(\Rs=1)$ diverges; therefore
    only the unfolded phase exists for $\Rc\geq\Rcs$.  For $\Rc\leq2$, there is no melting
    transition at zero force as $\Rwc(\Rs=1) = 0$. Thus if no force is applied, the system is
    always in the folded phase regardless of the temperature. The molecule can be denatured,
    though, by applying an external force even for $\Rc\leq2$, as can be seen from the phase
    boundary for $\Rc = 1.5$. The filled circles denote the thermal denaturation transition point
    $\Rwc(\Rs = 1)$ in the absence of an external force for
    $\Rc=2.1,\,2.16,\,2.3$.
    \subref{fig:7}b~Phase diagram in the $\Rw$-$\Rc$ plane. With zero force, $\Rs = 1$ (solid
    line), the weight~$\Rwc$ drops to zero for $\Rc \rightarrow2$ and diverges as
    $\Rc\rightarrow\Rcs\approx2.479$. For finite force, $\Rs>1$ (dashed line), a phase
   transition is possible, even for $\Rc\leq2$. 
    \subref{fig:7}c~Phase diagram in the $\RF$-$\RT$ plane for $\Rc =
    2.3$. Below the phase
    boundary the folded state is present, above the unfolded phase. Re-entrance at constant
    force is observed, as reported by
    M\"uller~\cite{Mueller2003}. %
  }
  \label{fig:7}
\end{figure}

\Eq~\eqref{eq:27} determines the phase diagram. In \fig~\ref{fig:7}a
we show the phase transition between the folded and unfolded states of
a homopolymeric RNA in the $\Rw$-$\Rs$ plane for a few different
values of the loop exponent~$\Rc$ that correspond to an ideal polymer,
$\Rc=3/2$, and values between $\Rc=2.1$ and $\Rc=2.3$ as they are
argued to be relevant for terminal and internal loops of varying
topology including the effects of self-avoidance.  Below the
transition lines in \fig~\ref{fig:7}a, the chain is in the unfolded
(extended) state, above the line in the folded (compact) state. With
growing loop exponent, the extent of the folded phase shrinks. In
fact, $\Rwc(\Rs,\Rc)$ diverges for $\Rc\rightarrow\Rcs$, where $\Rcs
\approx2.479$, \cf \eq~\eqref{eq:35}. Thus, for $\Rc\geq\Rcs$ only the
unfolded phase is present, as observed earlier in the restricted case
of zero force, $\Rs=1$~\cite{Einert2008}.  On the other hand, the
critical line for $\Rc\rightarrow2$ and $\Rs=1$ goes down to zero,
$\Rwc(\Rs=1)\rightarrow0$ for $\Rc\rightarrow2$, which indicates that
if no external force is applied to the molecule, there is only the
folded phase for $\Rc\leq2$ and hence no thermal phase transition is
possible. However, applying a sufficient force can drive the molecule
into the unfolded state, as derived earlier.  Concluding, only for
$2<\Rc<\Rcs$ a thermal phase transition, denoted by the filled circles
in \fig~\ref{fig:7}a, is possible. A force induced phase transition is
possible whenever $\Rc<\Rcs$ and $\Rw>\Rwc$.  For small force and
$\Rc>2$, \eq~\eqref{eq:27} can be expanded around $\Rs = 1$ and yields
the universal asymptotic locus of the phase transition
\begin{equation}
  \label{eq:37}
  \Rwc(\Rs) \sim \Rwc(\Rs = 1) + (1-\Rs)^{\Rc-2}\frac{\Gamma(2-\Rc)}{(\RiemannZeta{\Rc-1}-2
    \RiemannZeta{\Rc})^2}
  \eqspace.
\end{equation}

In \fig~\ref{fig:7}b we show the critical line $\Rwc(\Rs,\Rc)$ for two
different values of $\Rs$ as a function of the loop exponent $\Rc$. In
the absence of an external pulling force, \ie for $\Rs=1$ (solid
line), the transition line only occurs in the limited range $2 < \Rc <
\Rcs\approx2.479 $. It is seen that for loop exponents around the
relevant value of $\Rc\approx 2.1$, the critical base pairing weight
is quite small and of the order of $\Rwc \approx 0.1$. A base pairing
weight smaller than unity corresponds to a repulsive base pairing free
energy that is unfavorable. This at first sight paradoxical result,
which means that the folded phase forms even when the extensive part
of the base pairing free energy is repulsive, reflects the fact that
the folded state contains a lot of topological entropy because of the
degeneracy of different secondary structures.
The consequences for the theoretical description of systems, where
single stranded nucleic acids occur, including DNA transcription,
denaturation bubbles in dsDNA, untwisting of nucleic
acids~\cite{Leger1999}, and translocation~\cite{Bundschuh2005} will be
briefly discussed in
section~\ref{sec:dna-melting}.  

The phase diagram, \eq~\eqref{eq:27}, can also be displayed in the
$\RF$-$\RT$ plane by virtue of \eqs~\eqref{eq:1} and~\eqref{eq:6} and
is shown in \fig~\ref{fig:7}c.  Here, re-entrance at constant force
becomes visible, in line with previous predictions by
M\"uller~\cite{Mueller2003}. %
Expanding \eq~\eqref{eq:6} we obtain $\Rs(\RF) \sim
1+(\RKuhnss\RF/(\kBT))^2/6$, for $\RF\rightarrow0$.  \Eq~\eqref{eq:37}
yields the scaling of the critical force close to the melting
temperature as
\begin{equation}
  \RFc \propto (\RTmelt - \RT)^{1/(2c-4)}\label{eq:38}
  \eqspace,
\end{equation}
which depends on the loop exponent~$\Rc$ and deviates from the
predictions by
M\"uller~\cite{Mueller2003}, %
who found a universal exponent $1/2$.

\subsection{Thermodynamic quantities and critical exponents}
\label{sec:therm-quant}
We now consider the thermodynamic and critical behavior of various
quantities.  An arbitrary extensive quantity $Y$ with the conjugate
field $f$ is obtained from the grand potential~$\RPG = -\kBT\ln\RZG$
\via differentiation with respect to $f$ and the chemical potential
$\Rmu$ held constant
\begin{equation}
  \label{eq:39}
  Y = \left.\frac{\partial \RPG}{\partial f}\right|_{\Rmu}
  \eqspace.
\end{equation}
To evaluate the behavior of $Y$ in the thermodynamic limit,
$\RN\rightarrow\infty$, one sets $\Rmu \rightarrow\Rmud$, where
$\Rmud$ is defined as the chemical potential, at which $\RN(\Rmu) =
-{\partial\RPG}/{\partial\Rmu}$ diverges, \ie $\RN(\Rmu)
\rightarrow\infty$ for $\Rmu\rightarrow\Rmud$.  Another route to
obtain $Y$ is to conduct the calculation in the canonical ensemble,
\ie $\RN = \Tconst$, and to use the dominating
singularity~\cite{Einert2010}, where
\begin{equation}
  \label{eq:40}
  Y =\frac{\partial \RGG}{\partial f} = \kBT \RN \frac{\partial \ln\Rzd}{\partial f}
  \eqspace,
\end{equation}
see \eq~\eqref{eq:15}.  In fact, for $\RN\rightarrow\infty$
\eqs~\eqref{eq:39} and~\eqref{eq:40} are equivalent and $\Rmud$ is
associated with the dominating singularity of $\RZG(\Rz,\Rs)$, namely
$\Rzd = \exp(\Rmud/(\kBT))$, which will be shown now.  The Gibbs free
energy~$\RGG$ and the grand potential~$\RPG$ are related \via a
Legendre transform
\begin{equation}
  \label{eq:41}
  \RGG(\RN) = \RPG + \Rmu(\RN) \RN
  \eqspace.
\end{equation}
Therefore,
\begin{equation}
  \label{eq:44}
  Y
  =  \kBT\RN \frac{\partial \ln\Rzd}{\partial f}
  =\left.\frac{\partial \RGG}{\partial f}\right|_{\RN}
  =\left.\frac{\partial (\RGG-\RPG)}{\partial f}\right|_{\RN,\RPG}
  =N\left.\frac{\partial \Rmu}{\partial f}\right|_{\RPG}
  =\left.-\frac{\partial\RPG}{\partial\Rmu}\right|_{f}\left.\frac{\partial \Rmu}{\partial f}\right|_{\RPG}
  = \left.\frac{\partial \RPG}{\partial f}\right|_{\Rmu}
  \eqspace,
\end{equation}
where the two final expressions are evaluated at $\Rmu = \Rmud$.
While performing derivatives of the dominant singularity and of the
function $\Rk(\Rw,\Rz)$ is straightforward for $\Rzp$ and $\Rkp$, see
\eq~\eqref{eq:21}, one has to employ implicit differentiation of
\eq~\eqref{eq:16} to obtain the derivative of $\Rzb$ and $\Rkb$, see
supplementary material. For the latter case it turns out that
\eq~\eqref{eq:39} is more convenient to work with.

\subsubsection{Fraction of paired bases}
\label{sec:fraction-bound-bases}
\begin{figure}
  \centering
  \includegraphics{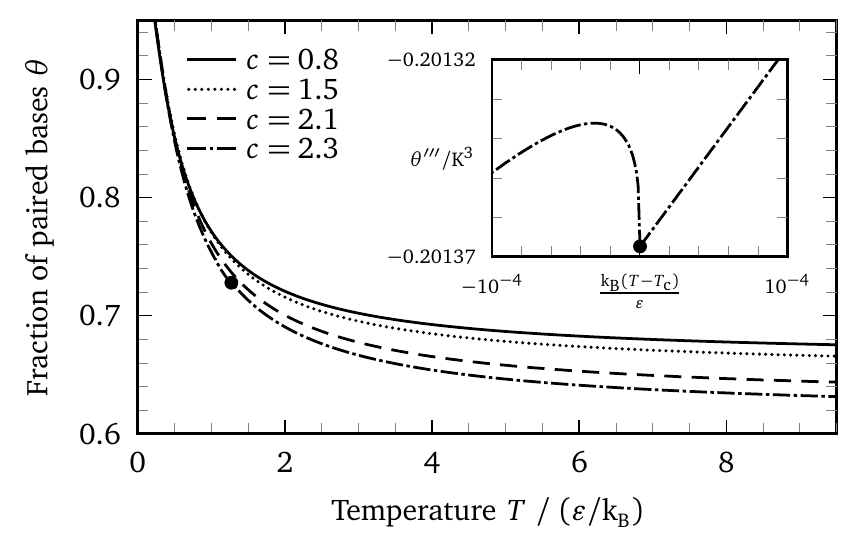}
  \caption{Fraction of paired bases as a function of temperature for
    $\Rw = \exp(\Reps/(\kBT))$ and various $\Rc =
    0.8,\,1.5,\,2.1,\,2.3$ at zero force, $\Rs = 1$. A phase
    transition is observed only for $\Rc =2.3$ for the range of
    positive values of $\Reps$ considered here (indicated by the
    filled circle), since for $\Rc \lesssim 2.195$ the critical weight
    of a hydrogen bond is $\Rwc < 1$, which can only be obtained for
    $\Reps<0$ amounting to a repulsive interaction. For $\Rc \leq 2$
    no thermal phase transition can be observed at all. The inset
    shows the third derivative $\Rnp'''=\drm^3\Rnp/\drm\RT^3$ for $\Rc
    = 2.3$, which reveals the phase transition.}
  \label{fig:8}
\end{figure}
\begin{figure}
  \centering
  \includegraphics{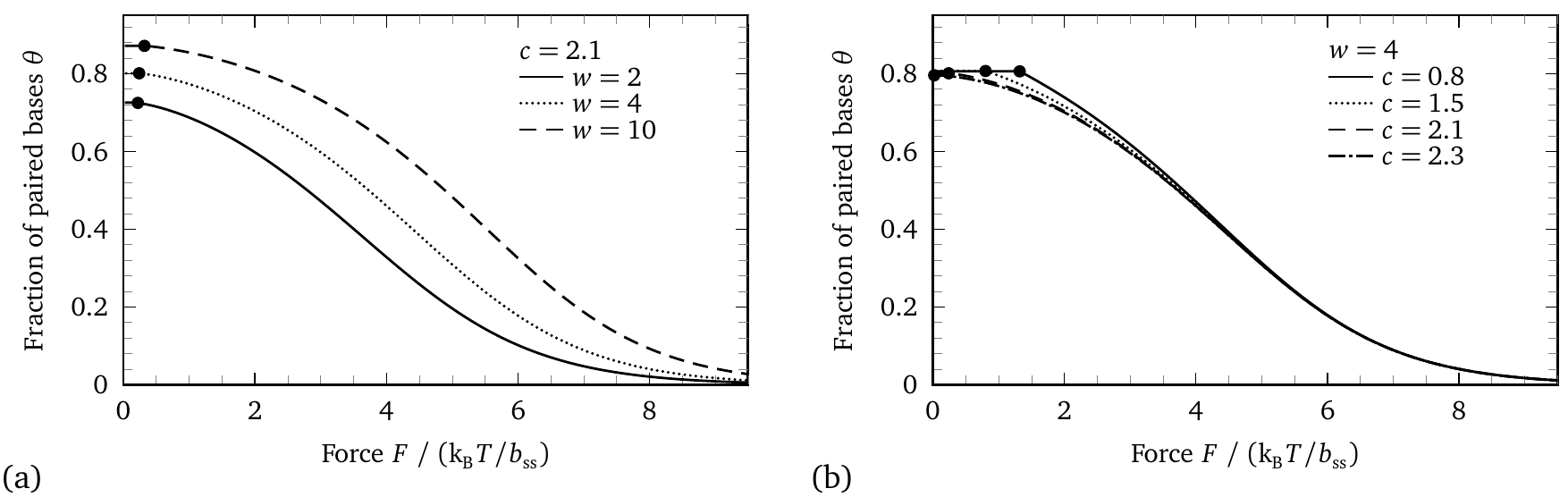}%
  \subfloat{\label{fig:9a}}%
  \subfloat{\label{fig:9b}}%
  \caption{Fraction of paired bases as a function of force for various
    $\Rw$ and $\Rc$. The phase transition is visible as a kink in the
    curves and indicated by a filled circle.  \subref{fig:9}a~$\Rc =
    2.1$ and varying $\Rw = 2,\,4,\,10$. \subref{fig:9}b~$\Rw = 4$ and
    varying $\Rc = 0.8,\,1.5,\,2.1,\,2.3$.}
  \label{fig:9}
\end{figure}
The fraction of paired bases is
\begin{equation}
  \label{eq:45}
  \Rnp = \frac{1}{\RN} \frac{\partial\ln\RQG_{\RN}}{\partial\ln\Rw}
  \eqspace.
\end{equation}
We obtain
\begin{equation}
  \label{eq:46}
  \Rnp = \frac{2\Li{\Rc}{\Rzb\Rk(\Rw,\Rzb)}}{\Li{\Rc-1}{\Rzb\Rk(\Rw,\Rzb)}}
\end{equation}
in the folded phase ($\RT<\RTc$, $\RF<\RFc$) and
\begin{equation}
  \label{eq:47}
  \Rnp = 1-\frac{1}{\sqrt{1+4\Rw\Li{\Rc}{1/\Rs}}}
\end{equation}
in the unfolded phase ($\RT>\RTc$, $\RF>\RFc$).  In \fig~\ref{fig:8}
the temperature dependence and in \fig~\ref{fig:9} the force
dependence of $\Rnp$ is shown.  The singularity at the critical point
of the thermal phase transition for zero force, $\Rs = 1$, is very
weak and becomes visible in the $\RPTn$\textsuperscript{th}
derivative, with $\RPTn$ being the integer with $(\Rc-2)^{-1} -1<\RPTn
< (\Rc-2)^{-1}$, see supplementary material.  The
$\RPTn$\textsuperscript{th} derivative exhibits a cusp, see inset of
\fig~\ref{fig:8},
\begin{equation}
  \label{eq:48}
  \frac{\drm^\RPTn\Rnp(\RT)}{\drm\RT^{\RPTn}} \propto |\RT - \RTmelt|^{\RnpexponentT} + \Tconst
  \eqspace,
\end{equation}
which is characterized by the critical exponent $\RnpexponentT =
(\Rc-2)^{-1} - \RPTn$ for $\RT < \RTmelt$ and $\RnpexponentT = 1$ for
$\RT > \RTmelt$, see supplementary material.  The force induced phase
transition is continuous, too, yet it exhibits a kink in $\Rnp(\RF)$
\begin{equation}
  \label{eq:49}
  \Rnp(\RF) \propto |\RF - \RFc|^{\RnpexponentF} + \Tconst
  \eqspace,
\end{equation}
which is characterized by the exponents $\RnpexponentF = 0$ for
$\RF<\RFc$ and $\RnpexponentF = 1$ for
$\RF>\RFc$. Therefore, the force induced phase transition
  is second order in accordance with previous results
  \cite{Mueller2003,Montanari2001}.  We note that for
$\RT\rightarrow\infty$, \ie $\Rw\rightarrow1$, a finite fraction of
bases are still paired. The situation is different for the force
induced phase transition where $\Rnp\rightarrow0$ for
$\RF\rightarrow\infty$.

\subsubsection{Specific heat}
\label{sec:specific-heat}
\begin{figure}
  \centering
  \includegraphics{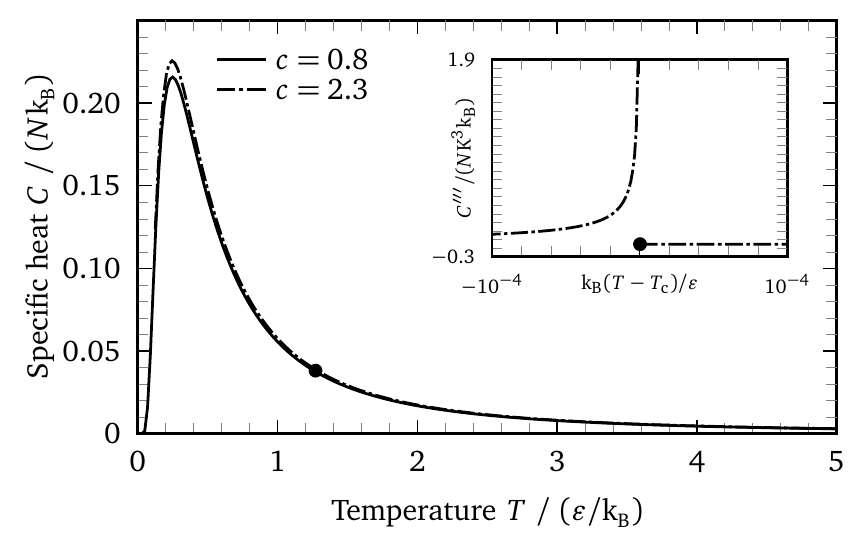}
  \caption{Specific heat as a function of temperature for different
    loop exponents $\Rc = 0.8,\,2.3$. The non-critical behavior of the
    heat capacity curve depends on the loop exponent only
    marginally. However, the existence and position of the critical
    point (indicated by the filled circle) and the critical behavior
    depend on $\Rc$. The inset depicts the third derivative of the
    specific heat, $\Rheatcap'''=\drm^3\Rheatcap/\drm \RT^3$, for $\Rc
    = 2.3$ revealing the phase transition.}
  \label{fig:10}
\end{figure}
The specific heat is defined as
\begin{equation}
  \label{eq:50}
  \Rheatcap = \frac{\kBT}{\RN}\frac{\partial^2 \RT\ln\RQG_{\RN}}{\partial \RT^2}
  \eqspace.
\end{equation}
One observes that the specific heat in \fig~\ref{fig:10} exhibits only
a very weak dependence on the loop exponent, which stands in marked
contrast to the findings for the short explicit sequence of
tRNA-phe~\cite{Einert2008,Einert2010a}, where a pronounced dependence
of the heat capacity on $\Rc$ is observed. The non-analyticity of
$\Rnp$, \eq~\eqref{eq:48}, translates into a divergence of the
$\RPTn$\textsuperscript{th} derivative of the specific heat at the
melting temperature
\begin{equation}
  \label{eq:51}
  \frac{\drm^\RPTn\Rheatcap(\RT)}{\drm\RT^{\RPTn}} \propto |\RT - \RTmelt|^{-\RheatcapexponentT} + \Tconst
  \eqspace,
\end{equation}
with the critical exponent $\RheatcapexponentT = \RPTn -
(3-\Rc)/(\Rc-2)$ for $\RT<\RTmelt$ and $\RheatcapexponentT = 1$ for
$\RT>\RTmelt$. This singularity is illustrated in the inset of
\fig~\ref{fig:10} for $\Rc = 2.3$.

\subsubsection{Fraction of non-nested backbone bonds}
\begin{figure}
  \centering
  \includegraphics{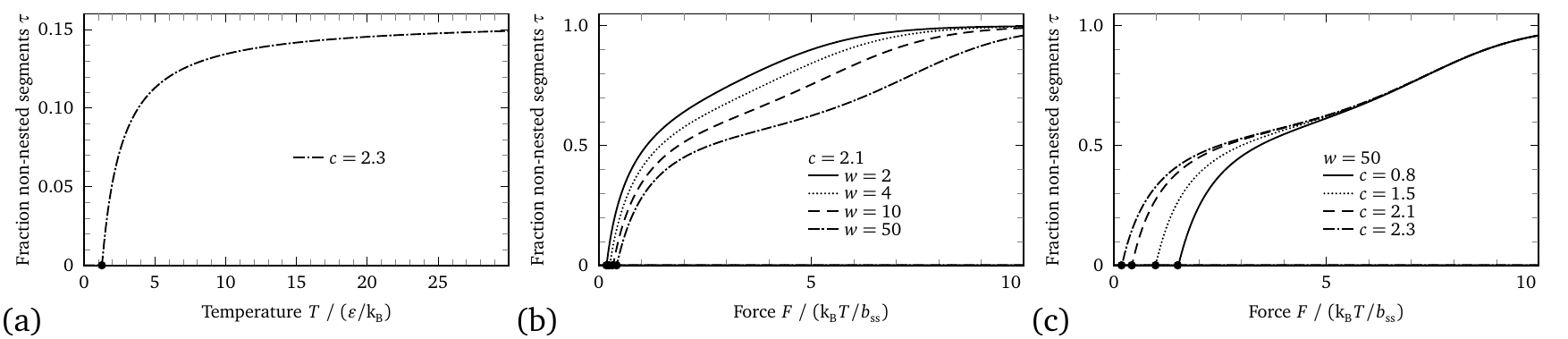}
  \subfloat{\label{fig:11a}}%
  \subfloat{\label{fig:11b}}%
  \subfloat{\label{fig:11c}}%
  \caption{\subref{fig:11}a~Fraction of non-nested backbone
    bonds~$\Rfm$ as a function of temperature for $\Rc = 2.3$. For
    $\Rc \lesssim 2.195$ the critical weight of a hydrogen bond is
    $\Rwc < 1$, which can only be obtained for $\Reps<0$, amounting to
    a repulsive interaction. Thus, for $\Rc \lesssim 2.195$ or
    $\RT<\RTmelt$ all segments are parts of loops or helices and hence
    $\Rfm = 0$. Filled circles indicate the melting temperature
    $\RTmelt$.  \subref{fig:11}b~and \subref{fig:11}c~fraction of
    non-nested backbone bonds as a function of force for various $\Rw$
    and $\Rc$. Again, the phase transition is visible as a kink in the
    curves and is indicated by a filled circle.  \subref{fig:11}b~$\Rc
    = 2.1$ and varying $\Rw = 2,\,4,\,10,\,50$. \subref{fig:11}c~$\Rw
    = 50$ and varying $\Rc = 0.8,\,1.5,\,2.1,\,2.3$. Filled circles
    indicate the position of the phase transition.}
  \label{fig:11}
\end{figure}\label{sec:fraction-non-nested}
The fraction of non-nested backbone bonds is obtained by
\begin{equation}
  \label{eq:52}
  \Rfm = \frac{1}{\RN}\frac{\partial\ln\RQG_{\RN}}{\partial\ln\Rs}
\end{equation}
and is
\begin{equation}
  \Rfm = 0\label{eq:57}
\end{equation}
in the folded phase, as $\Rzb$ does not depend on $\Rs$, and reads
\begin{equation}
  \label{eq:53}
  \Rfm
  = 1- \frac{2\Rw\Li{\Rc-1}{1/s}}{1+
    4\Rw\Li{\Rc}{1/s} + \sqrt{1+4\Rw\Li{\Rc}{1/s}}}
\end{equation}
in the unfolded phase.  For $\RT\rightarrow\infty$ the fraction of
non-nested backbone bonds assumes a finite value smaller than one,
which again indicates that the denatured phase in our model features
pronounced base pairing. However, for large force
$\RF\rightarrow\infty$ on invariably obtains $\Rfm\rightarrow1$.  As
can be seen nicely in \fig~\ref{fig:11}, both $\Rfm({\RT})$ and
$\Rfm({\RF})$ feature a kink at the critical point.

\subsubsection{Force extension curve}
\label{sec:force-extens-curve}
\begin{figure}
  \centering
  \includegraphics{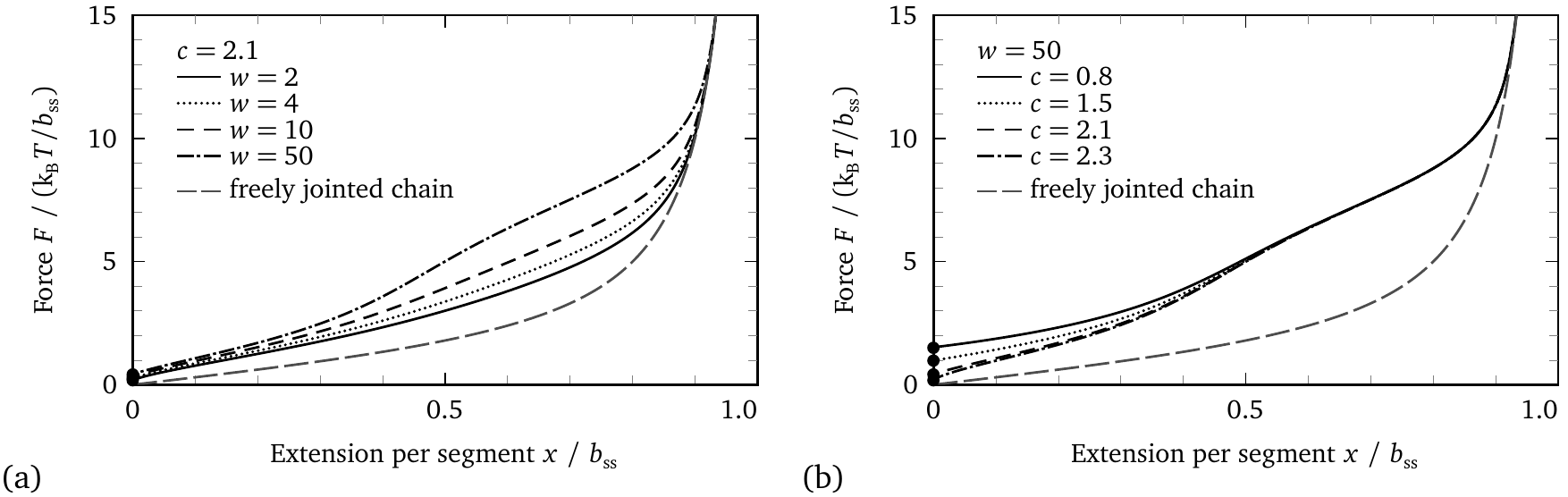}%
  \subfloat{\label{fig:12a}}%
  \subfloat{\label{fig:12b}}%
  \caption{Force extension curve as a function of force for various
    $\Rw$ and $\Rc$. The phase transition is indicated by filled
    circles and occurs at zero extension and a finite threshold force
    needed to unfold the compact folded structure.
    \subref{fig:12}a~$\Rc = 2.1$ and varying $\Rw =\exp(\Reps/\kBT)=
    2,\,4,\,10,\,50$. \subref{fig:12}b~$\Rw = 50$ and varying $\Rc =
    0.8,\,1.5,\,2.1,\,2.3$. Additionally the force extension curve of
    a freely jointed chain is plotted, which is the limiting form for
    $\Rw = 0$. }
  \label{fig:12}
\end{figure}
The force extension curve is closely related to the fraction of
non-nested backbone bonds $\Rfm$. The extension per monomer is given
by
\begin{equation}
  \label{eq:54}
  x(\RF) = \frac{\kBT}{\RN} \frac{\partial\ln\RQG_{\RN}}{\partial\RF} =
  \frac{\kBT}{\RN}\frac{\partial\ln\RQG_{\RN}}{\partial\ln\Rs}\frac{\partial\ln\Rs}{\partial\RF}
  = \RKuhnss \Rfm \timesspace(\coth(\beta\RF\RKuhnss)-1/(\beta\RF\RKuhnss)) = \RKuhnss\Rfm\timesspace\LangevinFunction(\beta\RF\RKuhnss)
  \eqspace,
\end{equation}
Since the Langevin function $\LangevinFunction$ is a smooth function,
the critical behavior of $x(\RF)$ is governed by the behavior of the
fraction of non-nested bonds $\Rfm$.  As can be seen in
\fig~\ref{fig:12}a, the stretching behavior of the Langevin function
is approached as the base pairing weight decreases, otherwise
pronounced deviations are seen in the force-stretching curves.  Also,
a finite stretching force to unravel the folded state is
needed. From \fig~\ref{fig:12} it becomes obvious that
  the force induced phase transition is second order as the force
  extension curve exhibits a kink at the critical force denoted by
  filled circles. The force extension curves are in accordance with
  previous results \cite{Montanari2001,Mueller2003}.

\section{Implications for DNA melting}
\label{sec:dna-melting}

How do the previous results impact on the theoretical description of
the denaturation of double stranded nucleic acid systems, particularly
DNA melting? When double stranded DNA approaches the denaturation
transition, more and more inter-strand base pairs break up and loops
proliferate.  In the traditional theories based on the Poland-Scheraga
model~\cite{Poland1966,Poland1966a}, the possibility of intra-strand
base pairing was not considered. These models are thus accurate for
duplexes formed between strands with sequences $[\mathrm{AG}]_{\RN/2}$
and $[\mathrm{TC}]_{\RN/2}$, where indeed base pairs (between A and T
and between G and C) can only form between the two strands, not within
one strand. For the case of duplexes formed by two strands with the
sequence $[\mathrm{AT}]_{\RN/2}$ or $[\mathrm{GC}]_{\RN/2}$, both
intra- and inter-strand base pairs can form and have identical
statistical weights. In this case, our model predicts significant
modifications for the duplex melting scenario.



The above sequence examples are prototypes for two extreme cases of
the general scenario characterized by statistical weights $\Rw$ and
$\Rwinter$ for intra-strand and inter-strand pairing,
respectively. Duplexes formed between $[\mathrm{AG}]_{\RN/2}$ and
$[\mathrm{TC}]_{\RN/2}$ are characterized by $\Rw = 0$, whereas
duplexes formed between two $[\mathrm{AT}]_{\RN/2}$ or
$[\mathrm{CG}]_{\RN/2}$ strands are characterized by $\Rw =
\Rwinter$. Intermediate values of $\Rw$ and $\Rwinter$ can be achieved
experimentally, for example, by $[\mathrm{ATT}]_{\RN/3}$ and its
complementary sequence $[\mathrm{AAT}]_{\RN/3}$, in which case only
$2/3$ of the intra-strand base pairs can be of the Watson-Crick type
and which would effectively lead to a lower weight of intra-strand
base pairs, \ie $\Rw<\Rwinter$. In naturally occurring DNA, a similar
situation might be present above the glass transition where, for
certain sequences, self-hybridization in a single strand is possible
to some extent~\cite{Bundschuh2002a}. The weights of inter-strand and
intra-strand base pairs can also be changed by applying an external
force or torque on the duplex, for instance in the setup by L\'eger
\emph{et al.}~\cite{Leger1999}.  We expect the weight $\Rwinter$ of
inter-strand base pairs to decrease when the duplex is untwisted. This
might lead to denatured regions in the duplex, where secondary
structure can form if the sequence allows. We speculate that
subsequent pulling first leads to the denaturation of the secondary
structures, whose signature would be a threshold force around
$\unit{1}{\pico\newton}$-$\unit{10}{\pico\newton}$
\cite{Montanari2001,Einert2010a}, see \fig~\ref{fig:12}, followed by
the over-stretching transition of DNA \cite{Einert2010}.  According to
our previous arguments, a marked dependence on the sequence should be
observed.


\begin{figure}
  \centering
  \includegraphics{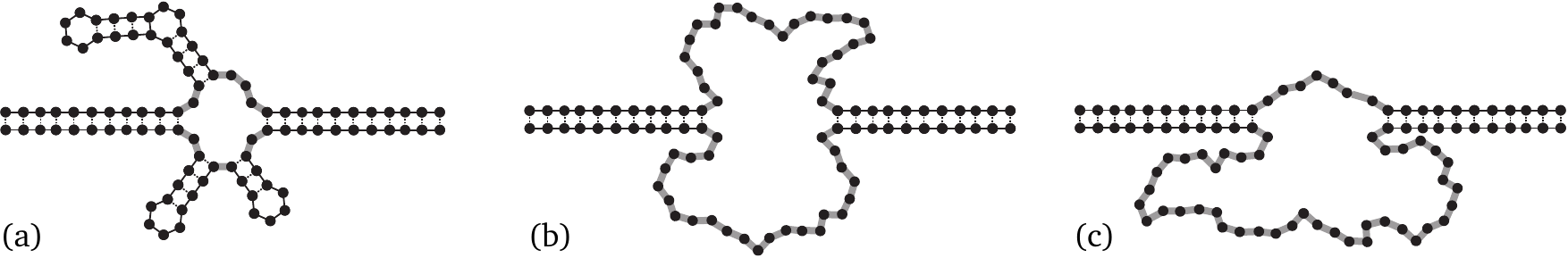}
  \caption{Illustration of a double stranded DNA molecule near the melting transition. (a)~If
    intra-strand base pairing is possible, in inter-strand loops the vast majority of bases
    will form intra-strand secondary structure elements. (b)~If the sequence does not favor
    intra-strand base pairs inter-strand loops without secondary structure will
    form. (c)~Illustration of an asymmetric loop.
  }
  \label{fig:13}
\end{figure}

Let us now review the classical Poland-Scheraga
calculation~\cite{Poland1966,Poland1966a} and allow --~in addition to
the standard treatment~-- for intra-strand base pairing in denatured
regions of the duplex, see \fig~\ref{fig:13}a.  The grand canonical
partition function of the two-state Poland-Scheraga model reads
\begin{equation}
  \label{eq:65}
  \DZPS = (1+\DZM) \sum_{k=0}^\infty(\DZB\DZM)^k(1+\DZB) -1 = \frac{\DZB + \DZM + 2 \DZB
    \DZM}{1-\DZB \DZM}
  \eqspace,
\end{equation}
where
\begin{equation}
  \label{eq:66}
  \DZB = \sum_{\RN=1}^\infty(\Rz\Rwinter)^\RN
  = \frac{\Rz\Rwinter}{1-\Rz\Rwinter}
\end{equation}
is the grand canonical partition function of a duplex in the bound
state with $\Rwinter$ the weight of an inter-strand base pair
and $\Rz$ now plays the role of the
fugacity of a base pair.
 We
distinguish between three different scenarios for a denatured or
molten region inside the double strand, which is characterized by the
partition function $\DZM$, depending on the sequence and the
intra-strand base pairing weight~$\Rw$: (i)~for $\Rw>\Rwc$, with
$\Rwc$ defined by \eq~\eqref{eq:27}, intra-strand secondary structures
in the folded (low temperature) phase are present, (ii)~for $\Rw<\Rwc$
intra-strand secondary structures in the unfolded (high temperature)
phase form, and (iii)~for $\Rw=0$ large denaturation bubbles occur
without intra-strand base pairing, corresponding to the original
Poland-Scheraga model. The grand canonical partition for case~(i) is
characterized by the branch point singularity, \eq~\eqref{eq:19},
\begin{equation}
  \label{eq:68}
  \DZM^{\mathrm{i}}= \sum_{\RN=1}^\infty \Rz^\RN\bigl(\Rzb^{-\RN} \RN^{-3/2}\bigr)^2 = \Li{\RcDNAeff}{\Rz/\Rzb^2}
  \eqspace,
\end{equation}
with $\RcDNAeff = 3$ and where the square is due to the fact, that on
either strand secondary structures may form; $\RN$-independent factors
have been neglected as they do not affect the critical behavior.
One realizes that the effective loop exponent~$\RcDNAeff$ in this case
is universal and independent of the configurational entropy of an
inter-strand loop characterized by the exponent~$\RcDNA$. For
case~(ii), the secondary structures on each strand are in the unfolded
phase and the partition functions are characterized by the pole,
\eq~\eqref{eq:24}.  As the number of non-nested segments, which is
proportional to $\RN$, is non-zero, \eq~\eqref{eq:53}, an inter-strand
loop decorated with helices occurs, and is characterized by the loop
exponent~$\RcDNA$. The respective grand canonical partition function
reads
\begin{equation}
  \label{eq:69}
  \DZM^{\mathrm{ii}}= \sum_{\RN=1}^\infty \Rz^\RN\bigl(\Rzp^{-\RN}\bigr)^2  \RN^{-\RcDNA} = \Li{\RcDNA}{\Rz/\Rzp^2}
  \eqspace.
\end{equation}
The third case constitutes the classical Poland-Scheraga model, where
no base pairing is present and the loop exponent $\RcDNA$ describes
the loop statistics. The partition function reads
\begin{equation}
  \label{eq:70}
  \DZM^{\mathrm{iii}}= \sum_{\RN=1}^\infty \Rz^\RN  \RN^{-\RcDNA} = \Li{\RcDNA}{\Rz}
  \eqspace.
\end{equation}
Combining \mbox{\eqs~(\ref{eq:65}-\ref{eq:70})} yields the grand
canonical partition function of a nucleic acid duplex
\begin{equation}
  \label{eq:74}
  \DZPS = \frac{\Rwinter \Rz + (1 + \Rwinter \Rz)
    \Li{\RcDNAeff}{\Rz/\RzbPS}}{1-\Rwinter\Rz(1+\Li{\RcDNAeff}{\Rz/\RzbPS})}
  \eqspace,
\end{equation}
with (i)~$\RcDNAeff=3$, $\RzbPS = \Rzb^2$, (ii)~$\RcDNAeff =
\RcDNA=2.1$, $\RzbPS = \Rzp^2$, or (iii)~$\RcDNAeff=\RcDNA=2.1$,
$\RzbPS = 1$, respectively.

\begin{figure}
  \centering
  \includegraphics{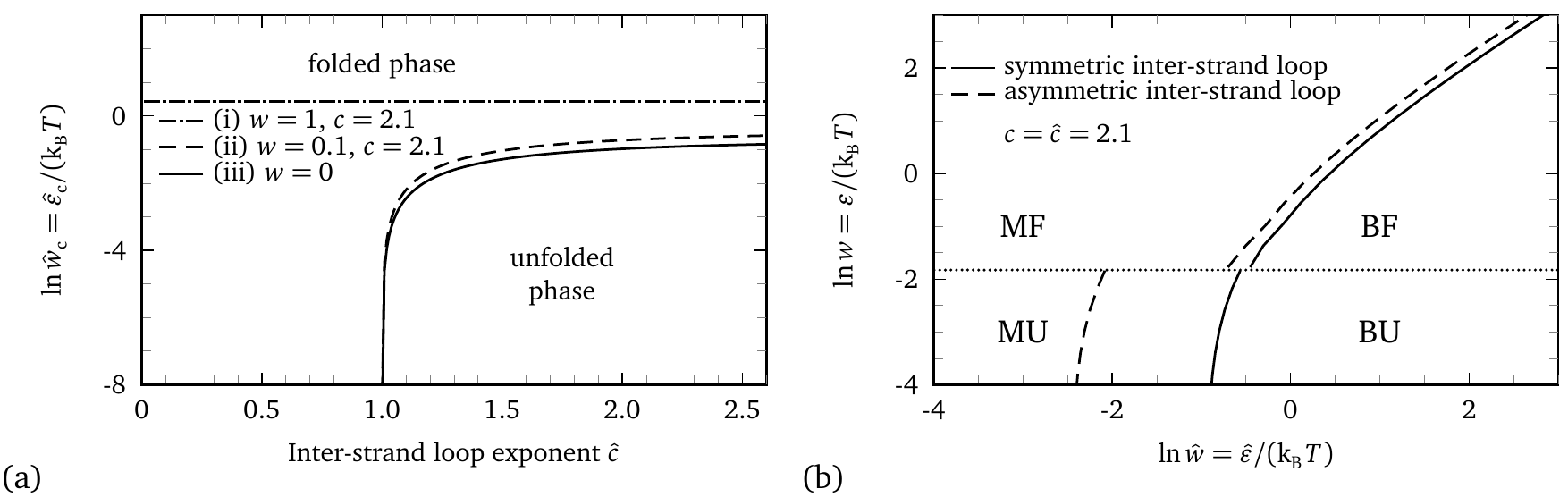}
  \caption{(a)~The critical inter-strand binding energy $\ln\RwcPS$ as
    a function of the inter-strand loop exponent~$\RcDNA$ for
    different intra-strand base pairing weights~$\Rw$ and different
    intra-strand loop exponents~$\Rc$. (i)~for $\Rw>\Rwc$, with
    $\Rwc(\Rc = 2.1) \approx 0.16$, see \eq~\eqref{eq:27}, inside
    denatured regions of the duplex secondary structures that are
    folded occur on each strand. (ii)~for $\Rw<\Rwc$ secondary
    structures that are unfolded form on each strand. (iii)~for
    $\Rw=0$ the classical Poland-Scheraga result is obtained, where no
    intra-strand base pairs form.  (b)~Phase diagram of a nucleic acid
    duplex in the plane spanned by the intra-strand pairing
    weight~$\Rw$ and the inter-strand pairing weight~$\Rwinter$ for
    symmetric (solid line) and asymmetric (dashed line) inter-strand
    loops for $\Rc =\RcDNA = 2.1$. Left to these curves the duplex is
    in the molten~(M) phase, to the right the duplex is in the
    bound~(B) phase.  The dotted line at $\Rw = \Rwc(\Rc = 2.1)
    \approx 0.16$, see \eq~\eqref{eq:27}, depicts the phase boundary
    between folded~(F) and unfolded~(U) intra-strand secondary
    structures for $\Rc = 2.1$, which separates cases~(i) and (ii),
    respectively. }
  \label{fig:14}
\end{figure}

The thermodynamics of nucleic acid duplexes is governed by the
singularities of the partition function, \eq~\eqref{eq:74}.  The
singularities are readily recognized as the branch point~$\RzbPS$ of
the polylogarithm~\cite{Abramowitz2002} and the pole~$\RzpPS$ of the
fraction in \eq~\eqref{eq:74}, which is the root of the denominator
\begin{equation}
  \label{eq:72}
  1 = \Rwinter\RzpPS(1+\Li{\RcDNAeff}{\RzpPS/\RzbPS})
  \eqspace.
\end{equation}
At the critical point, the pole and the branch point coincide, $\RzbPS
= \RzpPS$, which yields
\begin{equation}
  \label{eq:73}
  \RwcPS = \frac{1}{\RzbPS\bigl(1+\RiemannZeta{\RcDNAeff}\bigr)}
  \eqspace,
\end{equation}
where $\RiemannZeta{\RcDNAeff} = \Li{\RcDNAeff}{1}$ is the Riemann
zeta function.  In \fig~\ref{fig:14}a the critical inter-strand base
pairing weight~$\RwcPS$ as a function of the inter-strand loop
exponent~$\RcDNA$ is shown. The solid line depicts the phase diagram
of the classical Poland-Scheraga model~(iii) characterized by $\Rw =
0$, where no secondary structures occur in denatured regions, \cf
\fig~\ref{fig:13}b. The dashed line depicts the phase diagram for $\Rw
= 0.1$, where secondary structures occur that are in the unfolded
state since $\Rw=0.1 < \Rwc$ for $\Rc=2.1$, (ii). One sees
that the modifications to the $\Rw=0$ limit are rather small.
As in the case of the classical Poland-Scheraga model
there is no phase transition for $\RcDNA\leq1$, a second order phase
transition for $1<\RcDNA\leq2$, and a first order transition for
$\RcDNA>2$~\cite{Poland1966,Poland1966a,Kafri2000}. The situation is
different for case~(i) with $\Rw = 1$, where the secondary structures
are in the folded and the inter-strand loop exponent is replaced by
the universal value $\RcDNAeff = 3$, which renders the denaturation
transition first order and independent of $\RcDNA$. In
\fig~\ref{fig:14}b the phase boundary in the $\Rw$-$\Rwinter$ plane is
shown. The phase boundaries arising due to inter-strand base pairing
(solid or dashed line) and intra-strand base pairing (dotted line)
section the phase space into four quadrants, where the duplex is
either in the bound~(B) or molten~(M) state and the intra-strand
secondary structures are either folded~(F, case~(i)) or unfolded~(U,
case~(ii)).
Experimentally, a variation of temperature corresponds to a straight
path through the origin of the phase diagram in \fig~\ref{fig:14}b, which, depending on the values of the
inter- and intra-strand energies $\hat\Reps$ and $\Reps$, may cross a number of different phases.

There are $2\RN$ ways of constructing an asymmetric inter-strand
loop~\cite{Garel2004}, where the number of bases in the lower and the
upper part of a loop is not required to be identical, see
\fig~\ref{fig:13}c for an illustration. This additional factor reduces
the inter-strand loop exponent~$\RcDNA$ by~$1$. Therefore, the duplex
denaturation transition is for case~(i) right at the threshold between
continuous and discontinuous transitions, because $\RcDNAeff = 2$, and
for cases~(ii,iii) a continuous transition as $\RcDNAeff = 1.1$.
The consequences on the phase behavior are illustrated in \fig~\ref{fig:14}b
by broken lines.
We
add that the results for the competition between intra- and
inter-strand base pairing are obtained using a factorization
approximation for the two strands making up an inter-strand loop. In
the appendix we show that this factorization is accurate in the
thermodynamic limit of diverging loop size.





As a main result, the effective inter-strand loop exponent is
renormalized and takes on universal values for case~(i), where
intra-strand base pairing leads to secondary structures in the folded
phase.  The formation of intra-strand secondary structure influences
the melting temperature and has implications for the determination
base pairing free energy parameters \cite{SantaLucia1998,Xia1998} and
other biotechnological applications where DNA melting and
hybridization is involved. Thus, intra-strand interaction might be
important to include in software packages predicting the stability of
nucleic acids based on a Poland-Scheraga scheme. Algorithms for
cofolding of multiple nucleic acids already account for
this~\cite{Hofacker2003,Markham2005,Schuster2006}.

\section{Conclusions}
\label{sec:conclusions}
The partition function of RNA secondary structures has been evaluated
including arbitrary pairing topologies in the absence of pseudoknots,
including the configurational entropy of loops in the form of the loop
length dependent term $ \RGlentropy = - \kBT \ln \Rm^{-\Rc}$.  Exact
expressions for the fraction of paired bases, the heat capacity, and
the force extension curves are derived in the presence of an external
pulling force.  The observed thermal phase transition is very weak and
of higher order, the force induced transition is found to be second
order.  The critical behavior and the critical exponents are found to
depend on the loop exponent~$\Rc$.  A temperature induced melting
transition is only possible for $2<\Rc<\Rcs\approx2.479$.  Our theory
has consequences on the denaturation of double stranded DNA molecules,
in particular when intra-strand base pairs as well as inter-strand
base pairs can form. In this case, the native double strand is in
competition with intra-strand base pairing effecting the secondary
structures discussed in this paper.  Future directions will include
loop exponents, that depend on the number of helices emerging from a
given loop, treatment of pseudoknots, and cofolding nucleic acids to
study the influence of intra-strand interactions during double strand
denaturation.

\begin{acknowledgement}
  Financial support comes from the DFG \via grant NE 810/7.
  T.R.E. acknowledges support from the Elitenetzwerk Bayern within the
  framework of CompInt.
\end{acknowledgement}

\appendix

\section{Appendix}
\label{sec:appendix}

The statistical weight of a $\RN$ base pair long molten region in the
duplex with intra-strand interaction, as depicted in
\fig~\ref{fig:13}a, is given by
\begin{equation}
  \label{eq:55}
  \RQGDNA_{\RN} = \sum_{\RM, \RM'=0}^{\RN}\Rs^{\RM+\RM'+4} \frac{\RQHt_{\RN}^{\RM}
    \RQHt_{\RN}^{\RM'}}{(\RM+\RM'+4)^{\RcDNA}}
  \eqspace,
\end{equation}
where $\RcDNA=2.1$ is the loop exponent describing inter-strand loops
in DNA, $\RM+\RM'+4$ counts the number of non-nested back-bones that
contribute to the loop entropy, and $\RQHt_{\RN}^{\RM}$ is given by
\eq~\eqref{eq:4}. 
The denominator in \eq~\eqref{eq:55} is due to the loop entropy and amounts to an
effective interaction between the two strands as the expectation value
of $\RM$ depends on $\RM'$ and vice versa. The asymptotic behavior of
$\RQGDNA_{\RN}$ can be estimated by establishing two inequalities. The
first is
\begin{equation}
  \label{eq:60}
  \RQGDNA_{\RN} \leq \Rs^4 \sum_{\RM=0}^{\RN} \Rs^{\RM}\RQHt_{\RN}^{\RM} \sum_{\RM'=0}^{\RN}
  \Rs^{\RM'}\RQHt_{\RN}^{\RM'} = \Rs^4 \RQG_{\RN}\RQG_{\RN}
  \eqspace,
\end{equation}
where the scaling of $\RQG_{\RN}$ is given by \eq~\eqref{eq:19}
or~\eqref{eq:24} depending on whether the secondary structures are in
the folded or unfolded phase, respectively.  The upper scaling
boundary is obtained by factorizing \eq~\eqref{eq:55} and therefore
removing the effective interaction, but retaining the loop entropy for
each individual strand
\begin{equation}
  \label{eq:61}
  \RQGDNA_{\RN} \geq  \sum_{\RM=0}^{\RN} \frac{\Rs^{\RM+2}\RQHt_{\RN}^{\RM}}{(\RM+2)^{\RcDNA}}
  \sum_{\RM'=0}^{\RN} \frac{\Rs^{\RM'+2}\RQHt_{\RN}^{\RM'}}{(\RM'+2)^{\RcDNA}}
  \eqspace.
\end{equation}
The scaling of \eq~\eqref{eq:61} follows from the dominant singularity
analysis of the generating function
\begin{equation}
  \label{eq:62}
  \RZGDNA(\Rz,\Rs) = \Rs^2\Rz^2 \sum_{\RN = 0}^\infty \sum_{\RM=0}^{\RN} \Rs^{\RM}
  \Rz^{\RN}\frac{\RQHt_{\RN}^{\RM}}{(\RM+2)^{\RcDNA}}
  = \frac{1}{\Rk(\Rw,\Rz)} \Li{\RcDNA}{\Rs\Rz\Rk(\Rw,\Rz)} - \Rs\Rz
  \eqspace.
\end{equation}
$\RZGDNA(\Rz,\Rs)$ features the same singularities as $\RZG(\Rz,\Rs)$,
see \eq~\eqref{eq:9}, namely the branch point of $\Rzb$ of
$\Rk(\Rw,\Rz)$, and the singularity~$\Rzp$ given by the condition $1 =
\Rs\Rz\Rk(\Rw,\Rz)$, see \eq~\eqref{eq:21}.

By virtue of the inequalities~\eqref{eq:60} and \eqref{eq:61} we
conclude that
\begin{equation}
  \label{eq:6166}
  \RQGDNA_{\RN}   \propto \RzbPS^{-\RN} \RN^{-\RcDNAeff}
  \eqspace
\end{equation}
with $\RcDNAeff=3$, $\RzbPS = \Rzb^2$ for case (i) and $\RcDNAeff =
\RcDNA=2.1$, $\RzbPS = \Rzp^2$ for case (ii) for symmetric molten
loops.

\bibliographystyle{epj}
\bibliography{RNA_homo}

\clearpage



\section*{Supplementary material}

# Supplementary material
# Secondary structures of homopolymeric single-stranded nucleic acids including force and loop entropy: implications for hybridization

Thomas R. Einert[1], Henri Orland[2], and Roland R. Netz[1]

[1] Physik Department, Technische Universität München, James-Franck-Straße, 85748 Garching, Germany, Tel.: +49-89-28914337, Fax: +49-89-28914642, e-mail: einert@ph.tum.de
[2] Institut de Physique Théorique, CEA Saclay, 91191 Gif-sur-Yvette Cedex, France

April 28, 2011

**Abstract** This supplementary material contains explicit expressions for the derivatives of $\kappa$. The scaling behavior of the branch point $z_\mathrm{b}$ in the vicinity of the critical point is derived.

## Contents

## A  Derivatives of $\kappa$

In the main text we derived the three constitutive equations determining the thermodynamic behavior of the system

$$\kappa(w,z) = 1 + \frac{w}{\kappa(w,z)} \mathrm{Li}_c(z\kappa(w,z)) \tag{S1a}$$

$$\kappa(w, z_\mathrm{b}(w))^2 = w\mathrm{Li}_{c-1}(z_\mathrm{b}(w)\kappa(w, z_\mathrm{b}(w))) - w\mathrm{Li}_c(z_\mathrm{b}(w)\kappa(w, z_\mathrm{b}(w))) \tag{S1b}$$

$$sz_\mathrm{p}(w,s)\kappa(w, z_\mathrm{p}(w,s)) = 1 \; . \tag{S1c}$$

The function $\kappa(w,z)$ is the smallest positive root of eq. (S1a). Derivatives of $\kappa$ are obtained by implicit differentiation. To obtain the derivative of $\kappa$ with respect to an arbitrary quantity one differentiates both sides of eq. (S1a) with respect to this quantity and solves for the desired derivative. Explicitly,

$$\frac{\partial \kappa}{\partial z} = \frac{w\kappa \mathrm{Li}_{c-1}(z\kappa)}{z(\kappa^2 + w\mathrm{Li}_c(z\kappa) - w\mathrm{Li}_{c-1}(z\kappa))} \tag{S2}$$

$$\frac{\partial \kappa}{\partial w} = \frac{\kappa \mathrm{Li}_c(z\kappa)}{\kappa^2 + w\mathrm{Li}_c(z\kappa) - w\mathrm{Li}_{c-1}(z\kappa)} \tag{S3}$$

$$\frac{\partial \kappa}{\partial T} = -\frac{\kappa w}{T} \cdot \frac{\ln z \mathrm{Li}_{c-1}(z\kappa) + \ln w \mathrm{Li}_c(z\kappa)}{\kappa^2 + w\mathrm{Li}_c(z\kappa) - w\mathrm{Li}_{c-1}(z\kappa)} \; . \tag{S4}$$

## B Expansion of the branch point for $T < T_{\rm m}$

To obtain the critical behavior for $T < T_{\rm m}$ at zero force, $s = 1$, we perform an asymptotic expansion of eqs. (S1a) and (S1b) around the critical point, where the branch point and the pole coincide. Thus, at the critical point all eqs. (S1) have to hold and we obtain the critical values exactly as

$$\kappa_{\rm c} = \frac{1}{2}\left(1 + \sqrt{1 + 4w\zeta_c}\right) , \qquad z_{\rm c} = 2\left(1 + \sqrt{1 + 4w\zeta_c}\right)^{-1} , \quad \text{and} \quad w_{\rm c} = \frac{\zeta_{c-1} - \zeta_c}{(\zeta_{c-1} - 2\zeta_c)^2} . \tag{S5}$$

As an ansatz for $z_{\rm b}(T)$ and $\kappa_{\rm b}(T)$ we use a power series in $d = w/w_{\rm c} - 1$, where $w = \exp[\varepsilon/(k_{\rm B}T)]$ is the Boltzmann weight of a base pair. To simplify notation, we define $\alpha = (c-2)^{-1}$ and write

$$z_{\rm b}/z_{\rm c} \sim 1 + a_1 d + a_2 d^2 + \ldots + d^\alpha \left(a_\alpha + a_{\alpha+1} d + a_{\alpha+2} d^2 + \ldots\right) + a_{2\alpha-1} d^{2\alpha-1} + \ldots \tag{S6a}$$

$$\kappa_{\rm b}/\kappa_{\rm c} \sim 1 + b_1 d + b_2 d^2 + \ldots + d^\alpha \left(b_\alpha + b_{\alpha+1} d + b_{\alpha+2} d^2 + \ldots\right) + b_{2\alpha-1} d^{2\alpha-1} + \ldots \tag{S6b}$$

Plugging the ansatz (S6) into eqs. (S1a) and (S1b) we can solve order by order. To do so, the series representation of the polylogarithm $\mathrm{Li}_\nu(x)$ around $x = 1$ is used [1]

$$\mathrm{Li}_\nu(x) \sim \zeta_\nu - \zeta_{\nu-1}(1-x) + \frac{1}{2}(\zeta_{\nu-2} - \zeta_{\nu-1})(x-1)^2 + \ldots (1-x)^{\nu-1}\left(\Gamma(1-\nu) + \frac{1}{2}(1-\nu)\Gamma(1-\nu)(1-x) + \ldots\right) . \tag{S7}$$

We obtain the coefficients for $z_{\rm b}$

$$a_1 = -\frac{\zeta_c}{\zeta_{c-1}} \tag{S8a}$$

$$a_2 = \frac{\zeta_c^2}{\zeta_{c-1}^3}(2\zeta_{c-1} - \zeta_c) \tag{S8b}$$

$$a_3 = -\frac{\zeta_c^3}{\zeta_{c-1}^5}(5\zeta_{c-1}^2 - 6\zeta_{c-1}\zeta_c + 2\zeta_c^2) \tag{S8c}$$

$$a_4 = \frac{\zeta_c^4}{\zeta_{c-1}^7}(14\zeta_{c-1}^3 - 28\zeta_{c-1}^2\zeta_c + 20\zeta_{c-1}\zeta_c^2 - 5\zeta_c^3) \tag{S8d}$$

$$a_5 = -\frac{\zeta_c^5}{\zeta_{c-1}^9}(42\zeta_{c-1}^4 - 120\zeta_{c-1}^3\zeta_c + 135\zeta_{c-1}^2\zeta_c^2 - 70\zeta_{c-1}\zeta_c^3 + 14\zeta_c^4) \tag{S8e}$$

$$\vdots$$

$$a_\alpha = 0 \tag{S8f}$$

$$a_{\alpha+1} = -\frac{\Gamma(1-c) + \Gamma(2-c)}{\zeta_{c-1}}\left(-\frac{\zeta_{c-1}^2 - 3\zeta_{c-1}\zeta_c + 2\zeta_c^2}{\Gamma(2-c)\zeta_{c-1}}\right)^{\alpha+1} \tag{S8g}$$

$$a_{\alpha+2} = \frac{\left(\Gamma(1-c) + \Gamma(2-c)\right)\left(\zeta_{c-1}^2 + 4(c-2)\zeta_{c-1}\zeta_c + (5-3c)\zeta_c^2\right)}{(c-2)\zeta_{c-1}^3}\left(-\frac{\zeta_{c-1}^2 - 3\zeta_{c-1}\zeta_c + 2\zeta_c^2}{\Gamma(2-c)\zeta_{c-1}}\right)^{\alpha+1} \tag{S8h}$$

$$\vdots$$

$$a_{2\alpha-1} = 0 \tag{S8i}$$

$$\vdots$$

We obtain the coefficients for $\kappa_b$

$$b_1 = \frac{\zeta_c}{\zeta_{c-1}} \tag{S9a}$$

$$b_2 = -\frac{\zeta_c^2}{\zeta_{c-1}^3}(\zeta_{c-1} - \zeta_c) \tag{S9b}$$

$$b_3 = 2\frac{\zeta_c^3}{\zeta_{c-1}^5}(\zeta_{c-1} - \zeta_c)^2 \tag{S9c}$$

$$b_4 = -5\frac{\zeta_c^4}{\zeta_{c-1}^7}(\zeta_{c-1} - \zeta_c)^3 \tag{S9d}$$

$$b_5 = 14\frac{\zeta_c^5}{\zeta_{c-1}^9}(\zeta_{c-1} - \zeta_c)^4 \tag{S9e}$$

$$\vdots$$

$$b_\alpha = -\left(-\frac{\zeta_{c-1}^2 - 3\zeta_{c-1}\zeta_c + 2\zeta_c^2}{\Gamma(2-c)\zeta_{c-1}}\right)^\alpha \tag{S9f}$$

$$\begin{aligned} b_{\alpha+1} &= \frac{\Gamma(1-c) + \Gamma(2-c)}{\zeta_{c-1}}\left(-\frac{\zeta_{c-1}^2 - 3\zeta_{c-1}\zeta_c + 2\zeta_c^2}{\Gamma(2-c)\zeta_{c-1}}\right)^{\alpha+1} \\ &+ \left(-\frac{\zeta_{c-1}^2 - 3\zeta_{c-1}\zeta_c + 2\zeta_c^2}{\Gamma(2-c)\zeta_{c-1}}\right)^\alpha \frac{\zeta_{c-1}^2 - (c-2)\zeta_{c-1}\zeta_c - \zeta_c^2}{(c-2)\zeta_{c-1}^2} \end{aligned} \tag{S9g}$$

$$\begin{aligned} b_{\alpha+2} &= \frac{\zeta_{c-1} - \zeta_c}{2(c-2)^2 \cdot \Gamma(2-c)\zeta_{c-1}^4}\left(-\frac{\zeta_{c-1}^2 - 3\zeta_{c-1}\zeta_c + 2\zeta_c^2}{\Gamma(2-c)\zeta_{c-1}}\right)^\alpha \\ &\times \Big(2(c-2)\Gamma(1-c)(\zeta_{c-1} - 2\zeta_c)\big(\zeta_{c-1}^2 + 2(c-2)\zeta_{c-1}\zeta_c + (5-3c)\zeta_c^2\big) \\ &+ \Gamma(2-c)\big((c-3)\zeta_{c-1}^3 + (c-3)(4c-7)\zeta_{c-1}^2\zeta_c - (3c-5)(4c-9)\zeta_{c-1}\zeta_c^2 + (3c-5)(4c-7)\zeta_c^3\big)\Big) \end{aligned} \tag{S9h}$$

$$\vdots$$

$$b_{2\alpha-1} = -\frac{\zeta_{c-2} - 3\zeta_{c-1} + 2\zeta_c}{(c-2)\Gamma(2-c)}\left(-\frac{\zeta_{c-1}^2 - 3\zeta_{c-1}\zeta_c + 2\zeta_c^2}{\Gamma(2-c)\zeta_{c-1}}\right)^{2\alpha-1} \tag{S9i}$$

$$\vdots$$

Remarkably, the expansion of the product $z_b\kappa_b$ yields

$$z_b\kappa_b \sim 1 + (a_\alpha + b_\alpha)d^\alpha + (a_{\alpha+1} + b_1 a_\alpha + b_{\alpha+1} + a_1 b_\alpha)d^{\alpha+1} + \mathcal{O}(d^{\alpha+2}), \tag{S10}$$

meaning that all integer powers $d^n$, for $n < \alpha$, vanish. $n$ is the integer with $(c-2)^{-1} - 1 < n < (c-2)^{-1}$. This fact renders the transition of high order when $c \to 2$ and thus $\alpha \to \infty$. For the fraction of bound bases this can be seen directly by expanding the low temperature expression

$$\theta = 2\frac{\text{Li}_c(z_b\kappa_b)}{\text{Li}_{c-1}(z_b\kappa_b)} \tag{S11}$$

using eqs. (S7) and (S10).

# References

1. A. Erdélyi, *Higher Transcendental Functions*, Vol. 1 (McGraw-Hill, 1953)



\end{document}